\documentclass{article}
\usepackage{arxiv}
\usepackage[utf8]{inputenc}
\usepackage[T1]{fontenc}
\usepackage{microtype}
\usepackage{graphicx}
\usepackage{url}
\usepackage{booktabs}
\usepackage{amsmath, amssymb, amsfonts}
\usepackage{nicefrac}
\usepackage{xcolor}
\usepackage{caption}
\usepackage{algorithm}
\usepackage[noend]{algpseudocode}
\usepackage{array}
\usepackage{tabularx}
\usepackage{multirow}
\usepackage{longtable}
\usepackage{wrapfig}
\usepackage{tcolorbox}
\usepackage{tikz}
\usetikzlibrary{positioning,fit,backgrounds,shapes.geometric,arrows}
\usepackage{forest}
\usepackage{tocloft}
\usepackage{soul}
\usepackage{hyperref} % Move this to the end

\usepackage{soul}
\definecolor{lightgray}{gray}{0.7}
\setulcolor{lightgray}
\newcommand{\grayunderline}[1]{\setul{0.5ex}{0.3ex}\ul{#1}}

\title{Graph-based Fake Account Detection: A Survey}

\author{Ali Safarpoor Dehkordi \\
	School of Computing\\
	The Australian National University\\
	Canberra, Australia \\
	\texttt{ali.safarpoordehkordi@anu.edu.au} \\
	\And
	Ahad N. Zehmakan \\
	School of Computing\\
	The Australian National University\\
	Canberra, Australia \\
	\texttt{ahadn.zehmakan@anu.edu.au} \\
}

\newcommand{\shorttitle}{\@title}
\renewcommand{\shorttitle}{\textit{Graph-based Fake Account Detection: A Survey}}

\newcommand{\headeright}{}%A Preprint}
\newcommand{\undertitle}{}%A Preprint}

\begin{document}
\maketitle

\begin{abstract}
In recent years, there has been a growing effort to develop effective and efficient algorithms for fake account detection in online social networks.
This survey comprehensively reviews existing methods, with a focus on graph-based techniques that utilise topological features of social graphs (in addition to account information, such as their shared contents and profile data) to distinguish between fake and real accounts.
We provide several categorisations of these methods (for example, based on techniques used, input data, and detection time), discuss their strengths and limitations, and explain how these methods connect in the broader context.
We also investigate the available datasets, including both real-world data and synthesised models. We conclude the paper by proposing several potential avenues for future research.

\end{abstract}


\tableofcontents

\section{Introduction}
\label{sec:introduction}

Online Social Networks (OSNs), such as Facebook, Twitter (now X), and Instagram, have revolutionised how we interact, share information, and conduct business. These platforms have become essential tools for communication, entertainment, political discourse, and economic activities. Businesses leverage OSNs for marketing, customer engagement, and brand building, while individuals use them to express opinions, form communities, and stay informed about global events. Despite all their benefits, OSNs have also become a breeding ground for many destructive activities. They, particularly, facilitate the creation of fake accounts that can be used for various harmful purposes.

Whether labelled as counterfeit accounts, bots, sock puppets, fraudulent, or fake accounts, these fake personas pose significant risks to the security, trust, and integrity of online interactions.
Fake accounts are commonly used to manipulate public opinion~\cite{weng2022public}, spread misinformation, and commit financial crimes~\cite{link3statistics}. 
They exploit OSNs for political propaganda~\cite{ng2024assembling}, influence campaigns, and social engineering, undermining trust in digital spaces. Additionally, fake accounts contribute to artificial engagement~\cite{link3statistics}, inflating follower counts and interactions to distort perceived popularity, which can mislead accounts and businesses.

In this paper, we use the terms \textbf{fake} and \textbf{real accounts}. However, different studies adopt various terminologies. For example, fake accounts are also referred to as counterfeit accounts, deceptive accounts, bots, sybils, or fraudulent accounts. In contrast, real accounts are sometimes called legitimate accounts, genuine accounts, benigns, or authentic profiles.

Fake accounts have shown to be instrumental in influencing political events, such as the US $2020$ elections~\cite{ng2024assembling}, the Hong Kong $2021$ election~\cite{phillips2022competing}, and various electoral processes in Singapore $2020$, Indonesia $2019$, the Philippines $2019$, and Taiwan $2020$~\cite{uyheng2021active, uyheng2021computational}. Similar tactics have been observed in the French $2017$ presidential election~\cite{ferrara2017disinformation}, where fake accounts were leveraged for disinformation campaigns. Beyond politics, they have spread significant misinformation about COVID-$19$ and public health narratives~\cite{weng2022public, ng2024exploratory, ferrara2020types}.
% Furthermore, statistics show that fake accounts and scams on social networks have resulted in significant financial losses and security risks.
Fake engagements are reported~\cite{link3statistics} to cost advertisers $\$1.3$ billion in $2019$, mainly due to payments made to influencers with artificially inflated audiences.
From January to August $2024$, Australians lost $\$43.4$ million to social media scams, with $\$30$ million linked to fake investments~\cite{link1statistics}.
Furthermore, fake accounts continue to pose a growing cyber threat. For example, Facebook, Instagram, and Twitter removed $1.3$ billion, $95$ million, and $20$ million fake accounts in $2021$, respectively~\cite{link2statistics}.

Due to all these risks and threats, Fake Account Detection (FAD) is an essential area of research in cybersecurity, OSN analysis, and Artificial Intelligence (AI). Over the years, computer scientists have developed various techniques to automate the detection process, improving accuracy and scalability.

\subsection{What Are Different Categories of FAD Methods?}
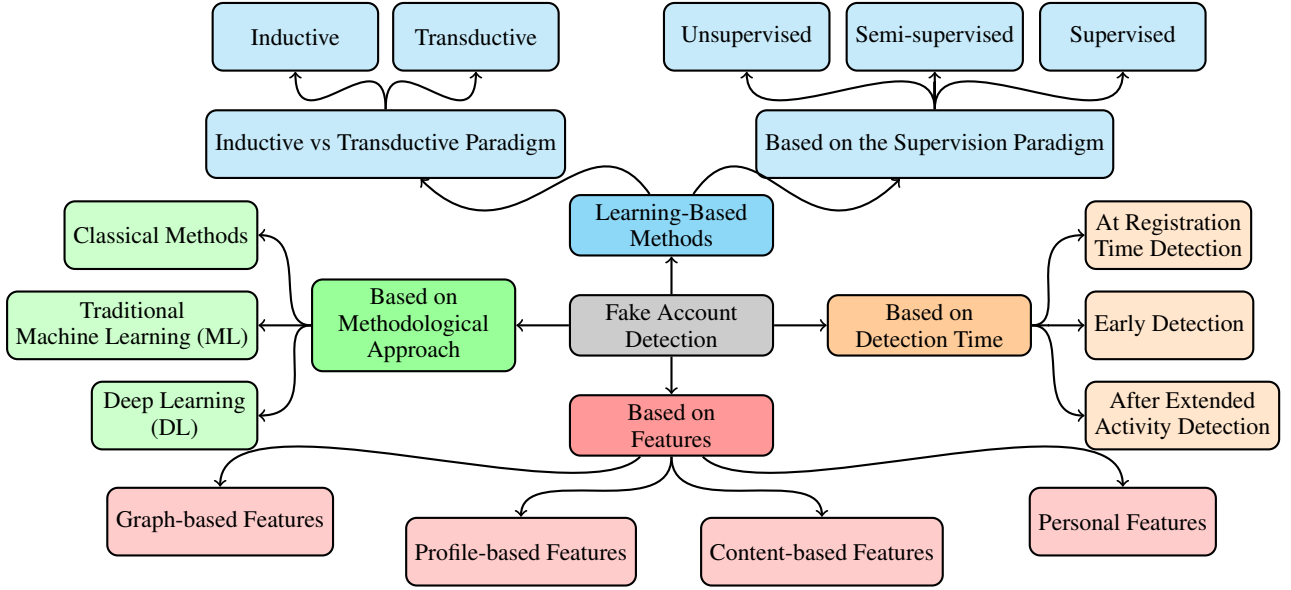
\begin{figure*}[ht]
\centering
\begin{tikzpicture}[
  node distance=.5cm and .7cm,
  every node/.style={draw, rounded corners, align=center, minimum width=2.7cm, minimum height=0.7cm, font=\small},
  root/.style={fill=gray!40},
  cat1/.style={fill=cyan!40},
  cat2/.style={fill=orange!40},
  cat3/.style={fill=red!40},
  cat4/.style={fill=green!40},
  leaf1/.style={fill=cyan!20},
  leaf2/.style={fill=orange!20},
  leaf3/.style={fill=red!20},
  leaf4/.style={fill=green!20},
  sub/.style={minimum width=2.2cm, minimum height=0.9cm},
  ->, thick
]

% Root node
\node[root] (root) {Fake Account\\Detection};

% First layer (categories)
\node[cat1, above=of root] (learn) {Learning-Based\\Methods};
\node[cat2, right=of root] (time) {Based on\\Detection Time};
\node[cat3, below=of root] (features) {Based on\\Features};
\node[cat4, left=of root] (method) {Based on\\Methodological\\Approach};

% Second layer - Learning Based
\node[sub, leaf1, above=of learn, xshift=-3.8cm, yshift = -0.3cm] (seen) {Inductive vs Transductive Paradigm};
\node[sub, leaf1, above=of learn, xshift=3.5cm, yshift = -0.3cm] (labels) {Based on the Supervision Paradigm};

\node[sub, leaf1, above=of seen, xshift=-1.2cm] (inductive) {Inductive};
\node[sub, leaf1, above=of seen, xshift=1.2cm] (transductive) {Transductive};

\node[sub, leaf1, above=of labels, xshift=-2.5cm] (unsup) {Unsupervised};
\node[sub, leaf1, above=of labels] (semisup) {Semi-supervised};
\node[sub, leaf1, above=of labels, xshift=2.5cm] (sup) {Supervised};

% Second layer - Time
\node[sub, leaf2, right=of time, yshift=1.2cm] (ext) {At Registration\\Time Detection};
\node[sub, leaf2, right=of time] (early) {Early Detection};
\node[sub, leaf2, right=of time, yshift=-1.2cm] (reg) {After Extended\\Activity Detection};

% Second layer - Features
\node[sub, leaf3, below=of features, xshift=-6cm, yshift = 0.1cm] (personal) {Graph-based Features};
\node[sub, leaf3, below=of features, xshift=-2cm, yshift = -0.3cm] (content) {Profile-based Features};
\node[sub, leaf3, below=of features, xshift=2cm, yshift = -0.3cm] (profile) {Content-based Features};
\node[sub, leaf3, below=of features, xshift=6cm, yshift = 0.1cm] (graph) {Personal Features};

% Second layer - Method
\node[sub, leaf4, left=of method, yshift=1.2cm] (dl) {Classical Methods};
\node[sub, leaf4, left=of method] (ml) {Traditional\\Machine Learning (ML)};
\node[sub, leaf4, left=of method, yshift=-1.2cm] (classic) {Deep Learning\\(DL)};

% Curved edges from root
\draw (root) -- (learn);
\draw (root) -- (time);
\draw (root) -- (features);
\draw (root) -- (method);

% Edges from learn
\draw (learn) .. controls +(305:-2cm) and +(135:-2.5cm) .. (seen);
\draw (learn) .. controls +(235:-2cm) and +(45:-2.5cm) .. (labels);

\draw (seen) .. controls +(-90:-1.5cm) and +(-90:1.5cm) .. (inductive);
\draw (seen) .. controls +(-90:-1.5cm) and +(-90:1.5cm) .. (transductive);

\draw (labels) .. controls +(-90:-1.5cm) and +(-90:1.5cm) .. (unsup);
\draw (labels) .. controls +(-90:-1.5cm) and +(-90:1.5cm) .. (semisup);
\draw (labels) .. controls +(-90:-1.5cm) and +(-90:1.5cm) .. (sup);

% Edges from time
\draw (time) .. controls +(180:-1.9cm) and +(-180:2cm) .. (ext);
\draw (time) .. controls +(180:-1.9cm) and +(-180:2cm) .. (early);
\draw (time) .. controls +(180:-1.9cm) and +(-180:2cm) .. (reg);

% Edges from features
\draw (features) .. controls +(-135:1.7cm) and +(90:1.8cm) .. (personal);
\draw (features) .. controls +(-90:1.3cm) and +(90:1.2cm) .. (content);
\draw (features) .. controls +(-90:1.3cm) and +(90:1.2cm) .. (profile);
\draw (features) .. controls +(-45:1.7cm) and +(90:1.8cm) .. (graph);

% Edges from method
\draw (method) .. controls +(-180:1.9cm) and +(180:-2cm) .. (dl);
\draw (method) .. controls +(-180:1.9cm) and +(180:-2cm) .. (ml);
\draw (method) .. controls +(-180:1.9cm) and +(180:-2cm) .. (classic);

\end{tikzpicture}
\caption{Taxonomy of FAD methods, categorised by methodological approaches, features, detection time, supervision paradigm, and inductive vs transductive paradigm.}
\label{fig:compact-fad-taxonomy}
\end{figure*}

FAD methods can be categorised in multiple ways. Some of the most important categorisations are visualised in Figure~\ref{fig:compact-fad-taxonomy} and are explained in more detail in the following.

\paragraph{1. Based on Methodological Approach.}
FAD methods can be broadly classified into three main categories, reflecting how techniques in this field have evolved over time.
\begin{itemize}
    \item \textit{Classical methods} use algorithmic or probabilistic techniques like belief propagation and random walks (e.g.~\cite{wang2017sybilscar}), operating directly on the data, especially the graph structure, without requiring any training.

    \item \textit{Traditional ML-based methods} make use of traditional ML algorithms (excluding DL models) applied to extracted features and the network structure. These techniques attempt to identify fake accounts by learning from patterns in the provided data.

    \item \textit{DL-based methods}, especially those leveraging Graph Neural Networks (GNNs), represent a more recent and powerful line of work. These models learn directly from the graph structure and available features, enabling them to model complex relation patterns among nodes and their contextual dependencies across the graph. Recent developments in this area incorporate elements such as multimodal data~\cite{ilias2024multimodal}, Reinforcement Learning (RL)~\cite{yang2023rosgas},  temporal information~\cite{he2024dynamicity}, and adversarial attacks~\cite{wang2023my}, highlighting the growing sophistication of modern FAD systems.
\end{itemize}
This categorisation forms the foundation for the structure of this survey, guiding our discussion of existing methods and emerging trends.

\begin{figure}[ht]
    \centering
    \includegraphics[width=0.7\linewidth]{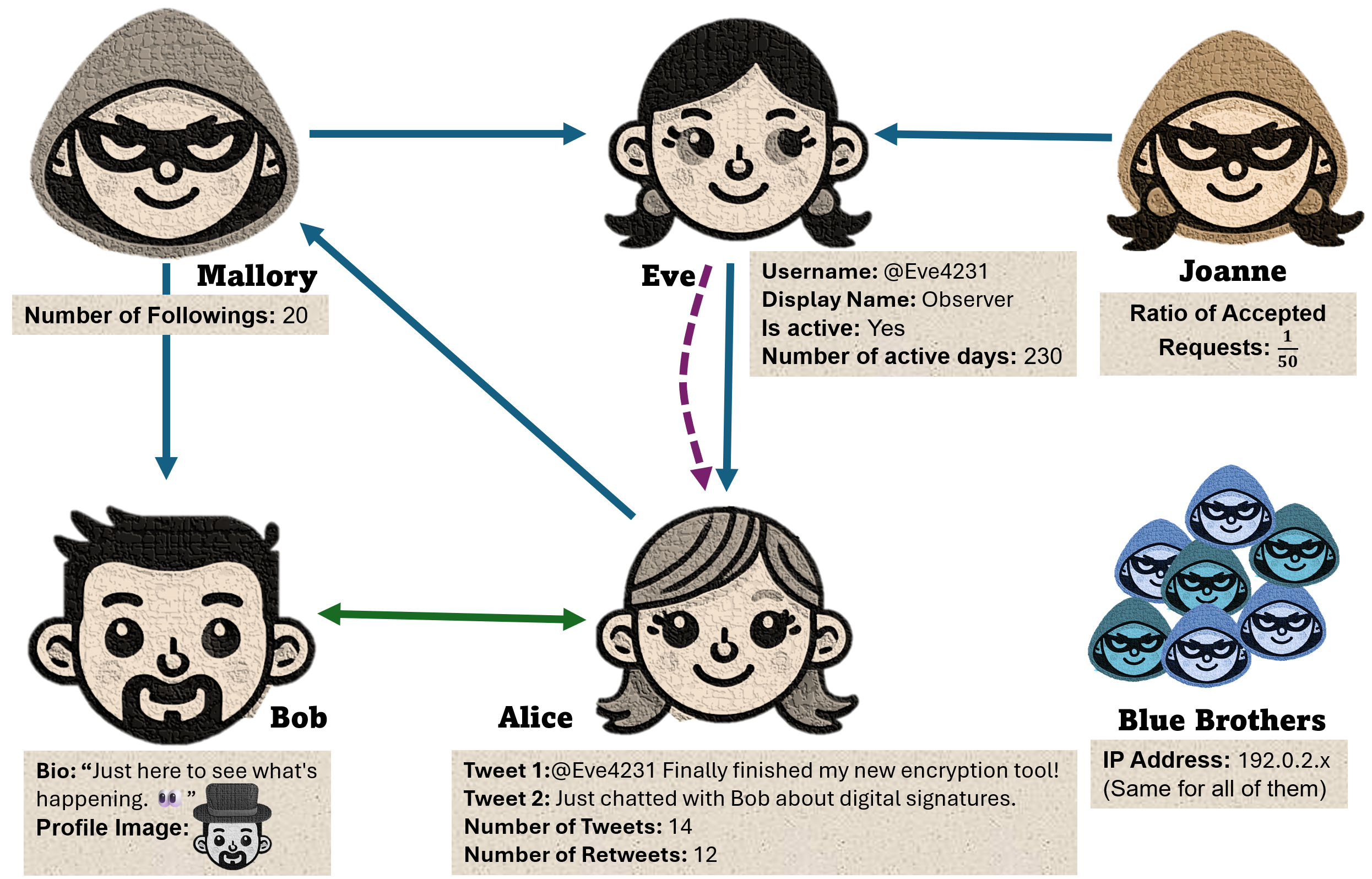}
    \caption{Different accounts with various features. Blue edges represent the \grayunderline{following relationships}, green edges indicate \grayunderline{mutual friendships}, and purple edges denote a \grayunderline{mentioned relationship}. Blue Brothers, Joanne, and Mallory are fake accounts that will be detected at different times.
    }
    \label{fig:introduction-FAD-CLS}
\end{figure}

\paragraph{2. Based on Features.}
FAD methods can also be distinguished based on the types of features they rely on for detection. In this survey, we categorise these features into the following groups. (Figure~\ref{fig:introduction-FAD-CLS} shows an example of different accounts with different features and connection types.)

\begin{itemize}
    \item \textit{Graph-based Features} are derived from the structure of the OSNs, such as node degrees~\cite{asghari2022using}, clustering coefficient, and Jaccard similarity~\cite{gao2015sybilframe}.
    % While many models operate directly on the graph structure, some particularly extract structural graph-based features to serve as input to these models.
    For example, in Figure~\ref{fig:introduction-FAD-CLS}, Mallory’s \grayunderline{Number of Followings} reflects her in-degree in the graph.
    
    \item \textit{Profile-based Features} include attributes explicitly listed in account profiles, such as usernames and account age (e.g.~\cite{boshmaf2015integro}). For example, in Figure~\ref{fig:introduction-FAD-CLS}, Eve has \grayunderline{Username}, \grayunderline{Display Name}, \grayunderline{Is Active}, and \grayunderline{Number of Active Days}, while Bob has a \grayunderline{Bio} and \grayunderline{Profile Image}.

    \item \textit{Content-based Features} capture the textual or multimedia content shared by accounts, like tweets or posts. In Figure~\ref{fig:introduction-FAD-CLS}, Alice’s \grayunderline{Tweets} and her \grayunderline{Number of Tweets and Retweets} exemplify content-based features.

    \item \textit{Personal Features} refer to sensitive account-specific attributes, such as IP addresses~\cite{liang2021unveiling}, or ratio of accepted requests~\cite{breuer2020sybiledge}. Although often considered part of profile features, we list them separately due to their specificity. For instance, in Figure~\ref{fig:introduction-FAD-CLS}, Joanne has \grayunderline{Ratio of Accepted Requests}, while the Blue Brothers share the identical \grayunderline{IP Address}.
\end{itemize}

This feature-based classification helps to clarify the assumptions and input requirements of different detection methods, shedding light on their applicability across various datasets and platforms.

In addition, a finer-grained categorisation can be made based on feature \textit{modalities}.
\textit{Numerical Features} represent scalar values, such as Mallory’s \grayunderline{Number of Followings} and Alice’s \grayunderline{Number of Tweets} in Figure~\ref{fig:introduction-FAD-CLS}.
\textit{Categorical Features} refer to discrete, non-numeric attributes, such as Eve’s \grayunderline{Is Active} status.
\textit{Textual Features} include account-generated text content such as Alice’s \grayunderline{Tweets} or Bob’s \grayunderline{Bio},
while \textit{Visual Features} include elements like Bob’s \grayunderline{Profile Image}. 
These modalities are not limited to the examples mentioned; for instance, the Blue Brothers’ \grayunderline{IP Address} represents a different kind of feature with a specific format. Furthermore, some features may change over time, and certain studies incorporate these temporal changes by treating them as \textit{dynamic features}(e.g.~\cite{zhang2023gufad}).

\paragraph{3. Based on Detection Time.}
FAD methods can also be categorised based on the detection time.
\begin{itemize}
    \item \textit{At Registration Time Detection} methods aim to identify the fake accounts early on, particularly during the registration time and before they join the network; this prevents potential harm before interacting with real accounts~\cite{liang2021unveiling}. 
    For example, in Figure~\ref{fig:introduction-FAD-CLS}, when the Blue Brothers register, they all use the same \grayunderline{IP Address}. This makes detection possible at the registration stage (cf.~\cite{xiao2015detecting}).
    \item \textit{Early Detection} means if an account slips through at the registration time detection stage, the next opportunity comes shortly after, based on its initial activity patterns. The goal is to flag fake accounts quickly before they blend in with the real ones (e.g.~\cite{sanchez2024early}). For instance, in Figure~\ref{fig:introduction-FAD-CLS}, while Joanne evades the first detection layer, she begins sending numerous requests to other accounts. However, as a fake account, most of her requests are rejected, resulting in a very low \grayunderline{Ratio of Accepted Requests}, and this behaviour can lead to early detection. Joanne still establishes a few connections in this case, but since she is detected early, she does not create many connections or share much content.
    \item \textit{After Extended Activity Detection} methods are designed to identify sophisticated fake accounts that evade detection mechanisms at the aforementioned two stages.
    At this stage, detection becomes more complex, relying on extended activity data to uncover fake accounts (e.g.~\cite{li2024botcl}). In our example, Mallory, having survived the second detection layer, successfully establishes numerous connections and begins sharing various types of content. At this stage, the third detection layer may successfully identify Mallory, considering his various graph and content features.
\end{itemize}

\begin{figure}[t]
    \centering
    \includegraphics[width=0.5\linewidth]{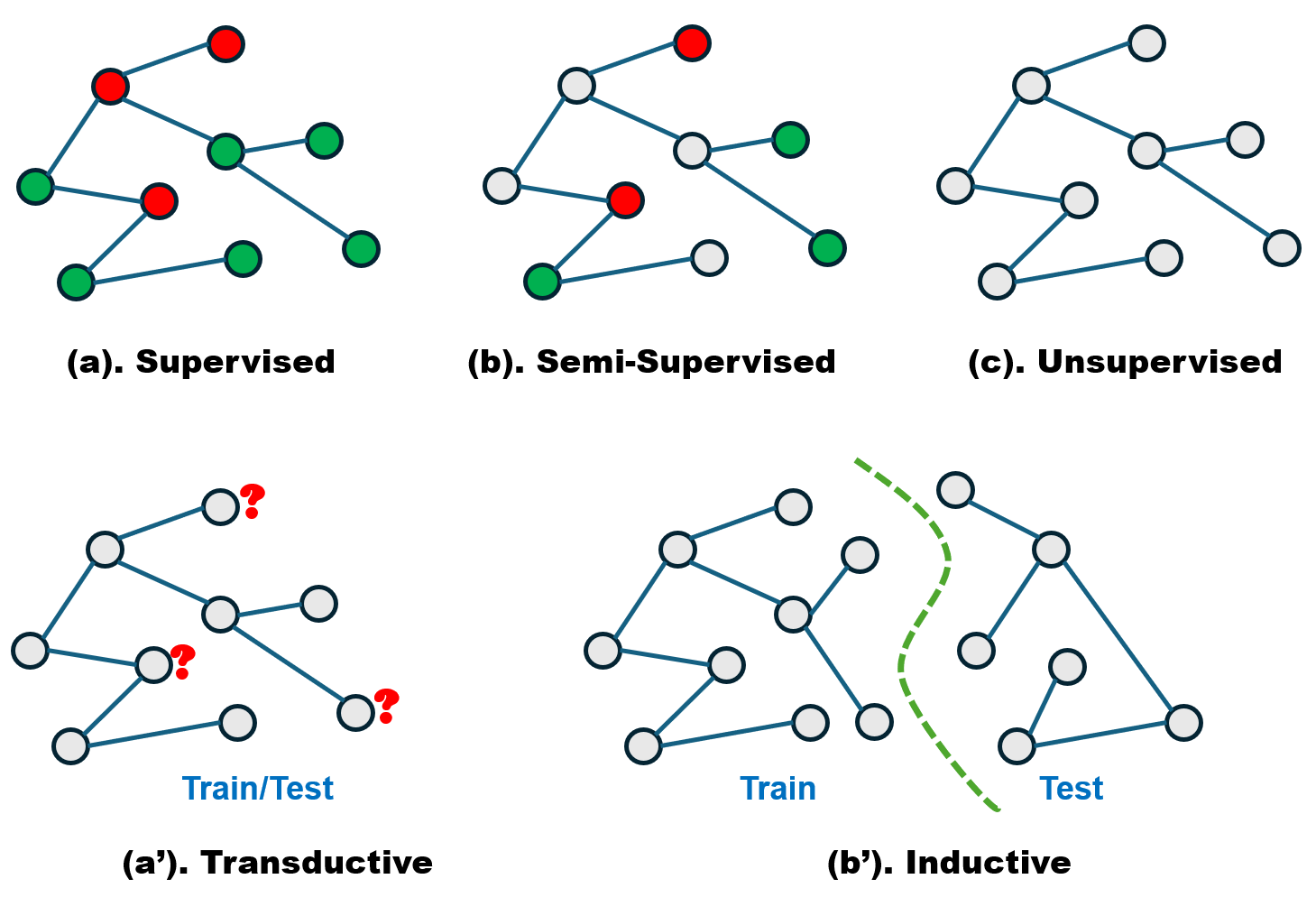}
    \caption{The first row represents Supervised (a), Semi-Supervised (b), and Unsupervised (c) scenarios, which are based on known labels/colours (in our case, real and fake). The second row illustrates transductive (a') and inductive (b') learning. The same graph is used in the transductive scenario, and test nodes are visible during training. In contrast, the inductive scenario involves two separate graphs (separated by a green line here).}
    \label{fig:introduction-model-cls}
\end{figure}

In \textbf{learning-based methods}, including both Machine Learning (ML) and Deep Learning (DL) approaches, we can further categorise these methods in the following two ways:

\paragraph{4. Based on the Supervision Paradigm.}
In ML and DL, FAD methods (generally, node classification methods) can be categorised based on how they utilise labels. \textit{Supervised} methods rely on labelled datasets to learn patterns that distinguish fake accounts from real ones. \textit{Semi-supervised} methods (e.g.~\cite{balaanand2019enhanced}) utilise a portion of labelled data while leveraging a larger pool of unlabelled instances. \textit{Unsupervised} methods (e.g.~\cite{peng2024unsupervised}) detect and classify similar behaviours without relying on labelled data.
In Figure~\ref{fig:introduction-model-cls}, (a), (b), and (c) represent supervised, semi-supervised, and unsupervised methods, respectively. In this illustration, nodes with known labels are coloured (red/green, corresponding to fake/real accounts in our case), while unlabelled nodes remain uncoloured.

\paragraph{5. Inductive vs Transductive Paradigm.}
Another important categorisation of FAD methods (and classification methods in general) is the split into transductive and inductive learning. Transductive methods (e.g.~\cite{peng2024coarse}) focus on predicting labels for nodes that are part of the training graph, using the same graph during both training and inference. In contrast, inductive methods (e.g.~\cite{heeb2024sybil}) generalise patterns to unseen data, which often involves using separate graphs for training and testing.
In Figure~\ref{fig:introduction-model-cls}, (a') and (b') correspond to transductive and inductive learning, respectively. In the transductive scenario (a'), the model has access to the entire graph during training and must infer the labels of nodes marked with a red question mark. In contrast, in the inductive scenario (b'), the training and test graphs are entirely separate.
It is worth noting that, in some studies, researchers create two separate subgraphs from the original graph for training and testing in the inductive setting. In other cases, they use entirely different graphs for training and testing.

\subsection{What Makes FAD So Challenging?}

Over time, fake account activities have become increasingly sophisticated.
Some fake accounts mimic real accounts to avoid detection, often exhibiting coordinated behaviours where multiple fake accounts collaborate to achieve fake account objectives~\cite{wang2023my}. Those responsible for managing fake accounts employ various techniques to manipulate network structures and evade detection. Some fake accounts form dense clusters to reinforce their presence, while others deliberately avoid clustering by rejecting connections with other fake accounts, making them appear more authentic. Some also attempt to blend into the network by connecting with as many real accounts as possible~\cite{yang2014uncovering}. Additionally, certain adversaries create fake accounts using real individuals' information~\cite{jethava2022user}, a tactic distinct from identity cloning attacks, where existing profiles are duplicated to impersonate victims~\cite{conti2012fakebook}.  
As FAD methods improve, fake account creators continuously refine their strategies. This leads to ongoing competition between detection systems and fake account creators. This dynamic results in an arms race in which detection methods must constantly adapt to new evasion techniques and sophisticated attack patterns.  

Advancements in generative AI and deepfake technology pose new challenges to FAD. Deepfake techniques generate highly realistic media, including images, videos, audio, and text, which can be used to create fake accounts, fabricate identities, spread misinformation, and manipulate public opinion. In the context of FAD, deepfake-generated content increases the credibility of fake information~\cite{srivasthav2024adaptive}, thereby aiding the fake accounts that utilise it and making them more difficult to distinguish from real ones.  
Fake accounts leveraging generative AI can create realistic profile pictures, bios, and posts, effectively bypassing traditional detection methods. To counteract this, advanced AI models are required to analyse inconsistencies, addressing the evolving nature of online deception~\cite{cresci2023demystifying, yang2024characteristics,ahmed2022analysis}.

Another important challenge in FAD is the limited availability of datasets. Determining whether an account is real or fake is typically a challenging task. Thus, the researchers usually resort to assumptions and approximate methods, which are not entirely accurate. For example,~\cite{wang2017sybilscar} assumes that accounts deleted after a certain period are fake, while those that remain active are real.
However, this assumption is flawed, as sophisticated fake accounts can persist for a long time, and many real accounts may be closed for reasons unrelated to being a fake account. As a result, many existing datasets often suffer from errors and inconsistencies, leading to noisy labels and potential misclassification. Moreover, the number of labelled samples is typically minimal, restricting the ability to train robust models.   
Class imbalance also presents a significant challenge in datasets, as the number of fake accounts can be much lower than that of real accounts. This imbalance increases the risk of biased outputs, making it difficult to achieve fair and reliable detection.
Privacy concerns also pose significant barriers to dataset availability, as ethical and legal restrictions limit access to account data. Ensuring compliance with data protection regulations further complicates the collection of high-quality, large-scale datasets for research purposes. Furthermore, many datasets contain incomplete graph structures, which limits their usefulness for graph-based methods. One key reason is the way these datasets are collected. As mentioned above, labelled accounts are often selected based on a specific sampling strategy. This can lead to labelled nodes being sparsely distributed across the network, meaning they are mostly far from each other. As a result, the available graph structure offers limited information for learning.

\subsection{What Is This Survey About?}

% Given the above limitation, fake accounts have limited control over their structural positioning within a network. Analysing graph properties provides a robust way to differentiate between real and fake accounts. For instance, 

As discussed above, a fake account, particularly one empowered with generative AI and deepfake technologies, can create profile and content information, such as name, posts, and comments, that closely resembles that of real accounts. Detecting such sophisticated fake accounts is highly challenging, especially for methods that investigate the account in isolation. While a fake account can easily manipulate such profile and content information, it has little to no control over its position with respect to the graph structure (especially the parts dominated by real accounts). This is because while a fake account can usually try to connect to another account (such as by sending a friend request), it is at the discretion of the receiving account to accept and form the connection. As a result, FAD methods that leverage graph structures have received a growing amount of attention in recent years.
Studies have demonstrated that leveraging graph structures significantly enhances detection accuracy (cf.~\cite{shi2023mgtab}).

In this survey, we explore FAD research that leverages graphs as the primary data source, emphasising the significance of graph-based approaches. Some studies integrate profile and content metadata alongside graph features, but others rely solely on graph structures for detection. While the survey is structured based on methodological approaches, including  Classical, Traditional ML, and CL methods, we also discuss other categories visualised in Figure~\ref{fig:compact-fad-taxonomy}, such as supervised/semi-supervised/unsupervised, inductive/transductive, and according to detection time, when appropriate.

\textbf{Outline.} The rest of this survey is organised as follows:
\begin{itemize}
    \item \textbf{Section~\ref{sec:preliminaries} (Preliminaries)} presents fundamental definitions, key terminology, and problem formulations. It introduces the graph-based representation of OSNs, the categorisation of fake and real accounts, and the concepts of attack edges, homophily, and heterophily.

    \item \textbf{Section~\ref{sec:classic_algos} (Classical Methods: Graph Algorithms and Probabilistic Mechanisms)} reviews non-learning-based approaches to FAD, including random walk-based methods (Section~\ref{subsec:cls:RW}), probabilistic inference and message passing methods such as loopy belief propagation (Section~\ref{subsec:cls:prop}), and structural preprocessing and feature integration methods (Section~\ref{subsec:cls:hyb}). It highlights key algorithms, their underlying assumptions (e.g., homophily), and their use of graph structure and local features for detection.

    \item \textbf{Section~\ref{sec:ML} (Traditional Machine Learning)} examines ML-based methods for FAD, excluding DL approaches. It covers feature-based models using classifiers such as Support Vector Machines (SVM) and Random Forest (RF), as well as graph-based ML methods (Section~\ref{subsec:ML-Graph}) and higher-order graph and probabilistic techniques (Section~\ref{subsec:ML-HigherOrder}).

     \item \textbf{Section~\ref{sec:DL} (Deep Learning)} presents graph-based DL approaches for FAD, generally focusing on how GNNs and their extensions improve detection by leveraging OSN structures. It begins with an overview of standard GNN-based methods and then discusses various advanced extensions: adaptations of classical methods (Section~\ref{subsec:DL:classic}), heterophily-aware architectures (Section~\ref{subsec:DL:heterophilic}), community and subgraph-level methods (Section~\ref{subsec:dl:subgraph}), multi-relational graphs (Section~\ref{subsec:DL-Het-MR}), content and multi-modal fusion techniques (Section~\ref{subsec:Incorporating Content Features}), contrastive learning (Section~\ref{subsec:DL-CL}), RL (Section~\ref{subsec:DL-RL}), temporal modelling (Section~\ref{subsec:DL-temporal}), mixture-of-experts frameworks (Section~\ref{subsec:DL-MOE}), federated learning (Section~\ref{subsec:federated_learning}), and adversarial attack resilience (Section~\ref{subsec:DL-AA}).

    \item \textbf{Section~\ref{sec:dataset} (Datasets)} describes commonly used datasets for FAD, including real-world OSN benchmarks (Section~\ref{subsec:dataset-benchmark}) and synthesised datasets (Section~\ref{subsec:dataset-synthesise}), highlighting their structural properties, annotation strategies, and relevance for evaluating detection methods.

    \item \textbf{Section~\ref{sec:conc+fut} (Conclusion and Future Work)} summarises key findings and outlines open research directions, including dataset development, heterophily handling, explainability, adversarial robustness, and learning under low supervision.
\end{itemize}

\section{Preliminaries}
\label{sec:preliminaries}

\paragraph{Graph Definitions.}
OSNs can be modelled as directed or undirected graphs $ G = (V, E) $, where accounts correspond to nodes, and their relationships form edges (both nodes and edges can be extended with additional features). In this graph representation, $ V $ is the set of nodes, and $ E \subseteq V \times V $ is the set of edges. An edge between two nodes $ u, v \in V $ is denoted as $ (u,v) \in E $ in a directed graph and $\{u,v\}$ in an undirected graph. We define $n = |V|$ and $m = |E|$, where $n$ and $m$ denote the total number of nodes and edges in the graph, respectively.
% We say a graph is undirected if for each edge $(v,u)\in E$, the edge $(u,v)$ is also in $E$.
For instance, account interactions, such as following another account, are generally represented by directed edges, while mutual relationships, such as friendships, are modelled as undirected (or bidirectional) edges.

The adjacency matrix is denoted by $ A \in \mathbb{R}^{n \times n} $. The entry $A_{i,j}$ is equal to the weight of edge $(i,j)$, which is denoted by $w_{i,j}$. If the graph is unweighted, we simply set $A_{i,j}=1$ if $(i,j)\in E$ and $0$ otherwise. For an undirected graph, the adjacency matrix is symmetric, meaning $ A_{i,j} = A_{j,i} $.

In an undirected graph, the set of neighbours is defined as $\Gamma(v) = \{ u \mid \{u,v\} \in E \}$ and the degree for node $v$ is defined as
$\deg(v) = |\Gamma(v)|$. 
In directed graph, for node $u$, set of in-neighbors is defined as $\Gamma_{\text{in}}(u) = \{ v \mid (v,u) \in E \}$ and set of out-neighbors define as $\Gamma_{\text{out}}(u) = \{ v \mid (u,v) \in E \}$.
In addition, for two subsets $V_1,V_2 \subseteq V$ we define $\partial(V_1, V_2) = \{(u,v) \in E: u \in V_1 \wedge v \in V_2 \}$.

We define the node feature matrix as $X \in \mathbb{R}^{n \times d}$, where each row $ x_v $ represents a $d$-dimensional feature vector associated with node $ v \in V $. The node embedding matrix is denoted as $ H \in \mathbb{R}^{n \times d'} $, where $ h_v $ is the learned $d'$-dimensional embedding for node $ v $. Node embeddings capture latent structural or semantic properties of nodes, distinguishing them from raw features represented by $ X $. 
Various research efforts (cf.~\cite{ali2019detect, feng2021satar, pham2022bot2vec}), broadly categorised under representation learning, seek to define or learn a function $ f_{\text{rep}}(\cdot) $ that generates a node’s embedding based on its features, i.e., $ h_{v_i} = f_{\text{rep}}(x_{v_i}) $. A common objective in representation learning is to achieve a lower-dimensional representation, typically with $ d' \ll d $ (cf.~\cite{zhu2020deep,khoshraftar2024survey}).

\paragraph{Fake and Real Account.}
In the context of FAD on a social graph $G$, we categorise accounts into two labels: \textit{fake} and \textit{real}. We denote the set of fake accounts as $F$ and the set of real accounts as $R$, so we have $V = F \cup R$ such that $F \cap R = \emptyset$. 
While each node is either fake or real, generally, in real-world datasets, the labels of some nodes are \textit{unknown}. We refer to these unknown nodes as $U$. Therefore, we define:
$V = F \cup R \cup U \text{ s.t. } (F \cap R) \cup (F \cap U) \cup (U \cap R) = \emptyset$. These sets are disjoint, meaning no node belongs to two or more categories. We define $\mathfrak{L}(v)$ as a function that returns a label of the node $v$. That is, for each node $v_i \in V$ we have $\mathfrak{L}(v_i) \in \{u, f, r\}$), where $u, f, r$ correspond to unknown, fake, and real.

\paragraph{Homophilic vs Heterophilic Edges.}{In this survey, we refer to \textit{homophilic edges} as edges that connect nodes with the same label (e.g., fake–fake or real–real), and \textit{heterophilic edges} as those that connect nodes with different labels (e.g., fake–real). Additionally, although the terms \textit{heterogeneous graph} or \textit{knowledge graph} are commonly used in prior research to address multiple node types or edge types in a graph, we adopt the term \textit{graph with multiple node or relation types} (also referred to as a \textit{multi-relational graph}) to avoid confusion with the label-based notion of heterophily.}

\paragraph{Fake and Real Region.}
We define the fake region as the subgraph induced by $F$, denoted as $G_F = (F, E_F)$, where $E_F \subseteq E$ represents the edges between fake accounts. Similarly, the real region is the subgraph induced by $R$, denoted as $G_R = (R, E_R)$, where $E_R \subseteq E$ represents the edges between real accounts. 

\paragraph{Attack Edges.}
Among edges in the network that represent account interactions, Attack edges are defined as the subset of edges connecting fake accounts to real accounts. Formally, the set of attack edges denoted as $E_A$, is defined as $E_A = \partial(F, R) = \{ (u,v) \in E \mid u \in F, v \in R\}$.
In this scenario, for an edge $ (u,v) \in E_A $, we refer to $ u $ as the \textit{attacker} and $ v $ as the \textit{victim}.
Since $F$ and $R$ are disjoint sets, attack edges always exist between nodes with different labels ($\forall (u,v)\in E_A: \mathfrak{L}(v) \ne \mathfrak{L}(u)$). These edges play an important role in analysing how fake accounts interact with real accounts, often aiming to manipulate, deceive, or influence the network. 

\begin{figure}[ht!]
    \centering
    \includegraphics[width=0.7\linewidth]{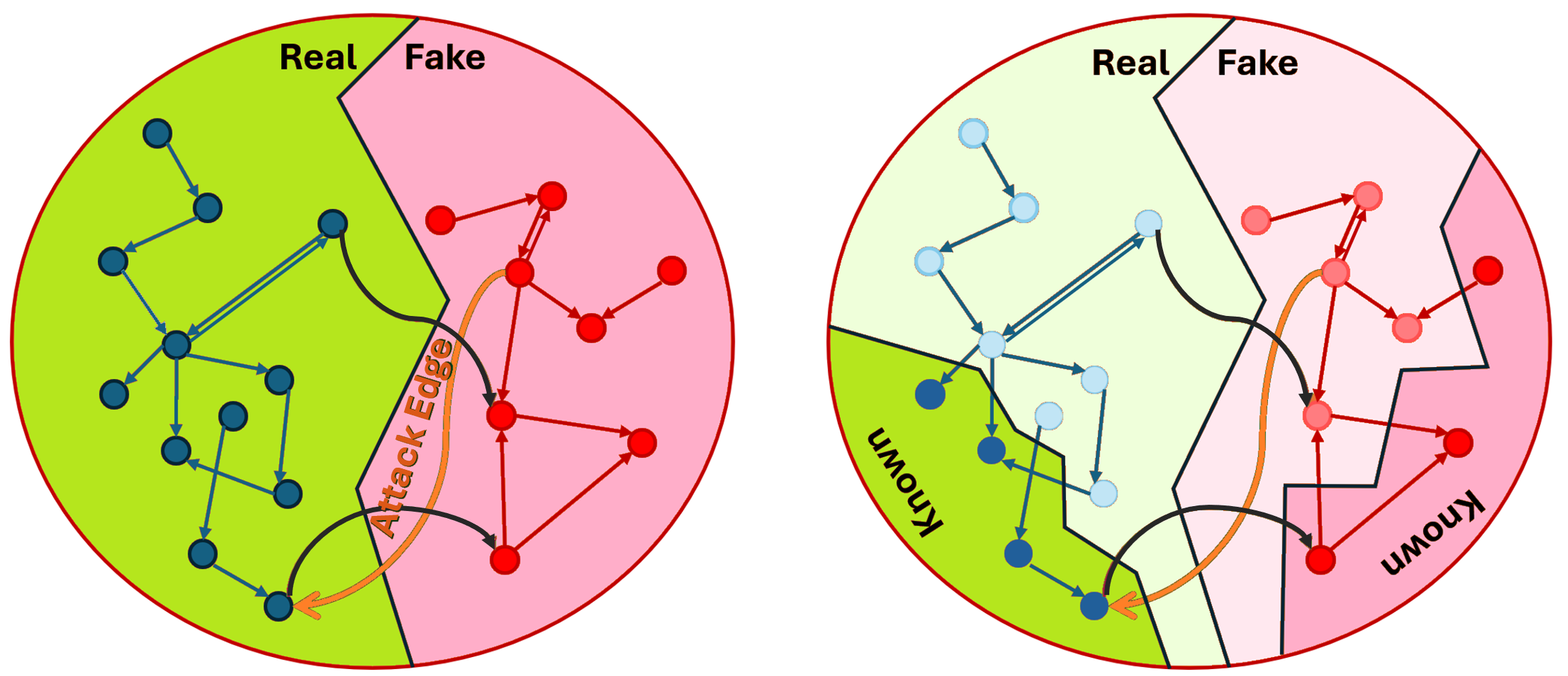}
    \caption{Structure of an OSN graph, illustrating real (green) and fake (pink) regions, homophilic edges (red and blue) and heterophilic edges (black and orange), and attack edges (orange) from fake to real nodes. The right side shows the same network with most nodes labelled as unknown, though they remain either fake or real (indicated by lighter shades).}
    \label{fig:graphsplit}
\end{figure}

Figure~\ref{fig:graphsplit} illustrates the structure of an OSN, divided into real and fake regions. Various types of edges are shown, including homophilic, heterophilic, directed, and attack edges. Additionally, in cases where node labels are unknown, they are represented using lighter shades for both nodes and regions on the right side of the figure.

\paragraph{Homophily/Heterophily in OSN.}
In graphs, homophily and heterophily are the tendencies or chance of nodes in an OSN to connect based on the similarity or dissimilarity of their labels~\cite{wu2023botscl}.
A network exhibits higher homophily (and lower heterophily) if nodes with the same label are more likely to be connected. 
Formally, homophily can be measured by the homophily ratio, defined as the proportion of edges connecting nodes with identical labels, as shown in Equation~\eqref{eq:homophily_heterophily}. Heterophily is defined as its complement. We say an edge $(u,v)$ is \textbf{homophilic} if $\mathfrak{L}(u) = \mathfrak{L}(v)$ and is \textbf{heterophilic edges} if $\mathfrak{L}(u) \ne \mathfrak{L}(v)$. 
\begin{equation}
    \textrm{homophily}(G = (V, E)) = \frac{1}{|E|} \sum_{(u,v) \in E} \mathbb{I}[\mathfrak{L}(u) = \mathfrak{L}(v)], \quad
    \textrm{heterophily}(G) = 1 - \textrm{homophily}(G)
    \label{eq:homophily_heterophily}
\end{equation}

In real-world datasets in the context of FAD, particularly in highly heterophilic networks, attack edges constitute a large proportion of the total heterophilic edges, as fake accounts primarily connect to real accounts to maximise their influence. In contrast, real accounts have less incentive to connect to fake accounts.

\paragraph{Homophily Assumption.}
Many FAD methods rely on the homophily assumption, which posits that real accounts predominantly connect with other real accounts, whereas fake accounts are often forced to form clusters among themselves, as their attempts to connect with real accounts are frequently rejected or limited~\cite{jia2017random, gong2014sybilbelief}.
Additionally, some studies introduce the small-cut assumption (e.g.~\cite{yu2006sybilguard}), which asserts that fake accounts have limited connections to real accounts, resulting in a small number of attack edges. In this assumption, the attack edges correspond to the small-cut in the graph.
Although these assumptions are framed differently, they convey the same fundamental principle: a structural separation exists between real and fake accounts, as the number of homophilic edges is much greater than the number of heterophilic edges. To maintain consistency, we refer to this overarching concept as the homophily assumption throughout this work.

Although many studies rely on the homophily assumption, some research has questioned its validity in real-world networks. Notably, Yang et al.~\cite{yang2014uncovering} conducted a large-scale empirical study on fake accounts in Renren, a Chinese OSN platform, and found that fake accounts are often embedded within real communities rather than isolated. Their study shows that coordinated behaviours can follow hidden patterns that are not easily captured through structural separation alone. This observation highlights the limitations of relying purely on the homophily assumption and motivates the development of more complex detection methods.

\paragraph{Features.}  
As mentioned in the Introduction (Section~\ref{sec:introduction}), FAD methods utilise various types of features to identify fake account behaviour. These features come in various forms: categorical (such as whether the account has a profile picture or is verified), numerical (like the number of followers or posts), textual (like usernames or the content the account shares), and visual (such as profile photos and shared images). We can also group features based on their origin. Profile features are taken directly from the account’s profile, such as the profile image and username. Content features originate from the account's shared content, such as tweets or videos. 
Graph-based features are derived from the structure of the social network graph, capturing how an account is connected to others. Examples include the number of connections (e.g., friends or followers) and structural metrics such as centrality or clustering coefficient.
Some features, referred to as personal features, include items such as IP addresses or phone numbers. These are typically private and not easily accessible.

\paragraph{Fake Account Detection (FAD).}
FAD is the task of identifying fake accounts within an OSN. The objective is to classify each account (node) as either \textit{fake} or \textit{real} by leveraging the network structure and any additional available features.
\vspace{1mm}

\begin{tcolorbox}[colback=blue!5,colframe=blue!45!black,title=FAD Problem Formulation]
\noindent 

\textbf{Input:}
An OSN is represented as a graph $G = (V, E)$; optional node features $X = \{x_v : v \in V\}$, and a set of labelled nodes partitioned into real accounts $R$ and fake accounts $F$ s.t. $R \cap F = \emptyset$.
The remaining nodes are unlabelled, i.e., $U = V \setminus (R \cup F)$. (In an unsupervised scenario, the labelled set is empty $R \cup F = \emptyset$.)

\noindent
\textbf{Output:} A classifier $f: V \to \{\text{real}, \text{fake}\}$
that assigns a label to each node in the graph.

\noindent
\textbf{Goal:} The primary objective is to maximise the correct identification of fake and real accounts. This is typically evaluated using performance metrics such as accuracy, precision, recall, and F1-score (see Table~\ref{tab:evaluationmetrics}).
\end{tcolorbox}

\begin{table*}[ht]
    \centering
    \renewcommand{\arraystretch}{2} 
    \resizebox{\textwidth}{!}{
    \begin{tabular}{|c|c|c|c|}
        \hline
        $TNR = \frac{TN}{TN+FP}$ & 
        $TPR = \frac{TP}{TP+FN}$ & 
        $FPR = \frac{FP}{TN+FP}$ & 
        $FNR = \frac{FN}{TP+FN}$ \\ 
        \hline
        \multicolumn{2}{|c|}{$\text{Accuracy} = \frac{TP + TN}{TP + TN + FP + FN}$} & 
        \multicolumn{2}{c|}{$\text{Balanced Accuracy} = \frac{1}{2} \left(TPR + TNR \right)$} \\ 
        \hline
        $\text{Precision} = \frac{TP}{TP + FP}$ & 
        $\text{Recall} = \frac{TP}{TP + FN}$ & 
        \multicolumn{2}{c|}{$F1 = 2 \cdot \frac{\text{Precision} \times \text{Recall}}{\text{Precision} + \text{Recall}}$} \\ 
        \hline
        \multicolumn{2}{|c|}{$\text{MCC} = \frac{(TP \times TN) - (FP \times FN)}{\sqrt{(TP+FP)(TP+FN)(TN+FP)(TN+FN)}}$} & 
        \multicolumn{2}{c|}{$\text{G-mean} = \sqrt{\frac{TP}{TP+FN} \times \frac{TN}{TN+FP}}$} \\ 
        \hline
        \multicolumn{4}{|c|}{$\text{Cohen's Kappa} = \frac{p_o - p_e}{1 - p_e}, \quad
            p_o = \frac{TP+TN}{TP+TN+FP+FN}, \quad
            p_e = \frac{(TP+FP)(TP+FN) + (TN+FP)(TN+FN)}{(TP+TN+FP+FN)^2}$} \\ 
        \hline
        \multicolumn{4}{|c|}{$\text{AUC} = \int_{0}^{1} TPR(FPR) \, dFPR 
        = \int_{0}^{1} TPR(FPR^{-1}(x))dx$} \\ 
        \hline
        \multicolumn{4}{|c|}{$\text{AUPRC} = \int_{0}^{1} \text{Precision}(\text{Recall}) \, d\text{Recall} 
        = \int_{0}^{1} \text{Precision}\left(\text{Recall}^{-1}(x)\right) \, dx$} \\ 
        \hline
    \end{tabular}
    }
    \caption{Common Evaluation Metrics}
    \label{tab:evaluationmetrics}
\end{table*}

\section{Classical Methods: Graph Algorithms and Probabilistic Mechanisms}
\label{sec:classic_algos}

In this section, we review classical (non-learning) algorithms, which can be broadly categorised into three principal groups based on their underlying methodologies.
First, Random Walk (RW)-based methods (Section~\ref{subsec:cls:RW}) leverage structural separation in the network, typically grounded in the homophily assumption.
Second, probabilistic inference and message passing methods (Section~\ref{subsec:cls:prop}) utilise probabilistic graphical models, particularly Loopy Belief Propagation (LBP), to iteratively refine node classifications based on both local priors and the global network structure.
Third, structural preprocessing and feature integration methods (Section~\ref{subsec:cls:hyb}) aim to improve detection performance by incorporating additional features or modifying the graph structure to facilitate more effective score propagation.

\subsection{Random Walk-Based Methods}
\label{subsec:cls:RW}

\paragraph{RW and Trace.}  The RW is a fundamental process in network analysis, in which a walker moves through a graph by transitioning from one node to a neighbouring node at each step according to a predefined probability distribution. In the simplest case, the walker selects a neighbour uniformly at random. That is, if the walker is currently at node $v$, the probability of moving to a specific neighbouring node is given by $ \frac{1}{\deg(v)} $, where $ \deg(v) $ denotes the degree of node $v$. More sophisticated RW variants incorporate edge weights, node centrality measures, or other structural properties to influence transition probabilities. In this survey, we use the term \textbf{trace} to refer to a single instance of an RW.
RWs are often used to measure how closely other nodes are connected to a specific node. By starting multiple traces from that node, the visit frequency of each node during traces indicates how strongly it is related to the starting node.

If nodes are initially assigned scores (e.g., prior estimated probability that a node is a fake account), RWs can be used to refine these scores. A straightforward method involves launching $r$ independent traces from each node $u$ and computing its score $p_u$ by aggregating the scores of visited nodes $v \in V$, weighted by the frequency of their visits.
While intuitive, this approach becomes computationally expensive on large-scale graphs, as it requires many traces per node to obtain accurate estimates. Here, the term score can vary in meaning depending on the context. For the FAD problem, it might represent the estimated probability that a node corresponds to a fake account.
Various papers use different terminology for this concept (e.g., trust score in~\cite{gao2018sybilfuse}). However, to maintain consistency and highlight the underlying similarity in behaviour, we use the term score as a general term.

To mitigate the inefficiency of the naïve method, an iterative approach is employed, where node scores are updated based on their neighbours’ scores rather than independently simulating multiple traces. This iterative formulation is inspired by well-established algorithms such as PageRank~\cite{page1999pagerank} and is widely used in score propagation-based methods like \textbf{SybilWalk}~\cite{jia2017random} and \textbf{SybilRank}~\cite{cao2012aiding}.
Let $ p_u^t $ denote the node $ u $ score at iteration $ t $. The simple update rule follows Equation~\eqref{eq:RWUpdateRule}. 

\begin{equation}
    p_u^t = \sum_{v \in \Gamma(u)} \frac{w_{u,v}}{\deg(v)} p_v^{t-1}
    \label{eq:RWUpdateRule}
\end{equation}

The iterative update can be expressed in matrix form for large-scale applications, facilitating efficient computation through linear algebraic operations. For instance, for the update rule provided in Equation~\eqref{eq:RWUpdateRule}, the transition matrix $ \mathbb{P} $ is defined as $\mathbb{P}[u,v] = \frac{w_{u,v}}{\deg(u)} $, and the score update can be written concisely as Equation~\eqref{eq:RWUpdateRuleMatrix}.
\begin{equation}
    p^t = \mathbb{P} p^{t-1}.
    \label{eq:RWUpdateRuleMatrix}
\end{equation}

An early approach to FAD, \textbf{SybilGuard}~\cite{yu2006sybilguard}, leveraged the homophily assumption, using constrained RWs to separate fake and real accounts. \textbf{SybilLimit}~\cite{yu2008sybillimit} improved efficiency by reducing the length of traces, making large-scale applications more feasible. Although these two methods were originally designed for P$2$P networks and were not directly applicable to OSNs, they influenced researchers to propose FADs based on homophily assumptions.

\textbf{SybilInfer}~\cite{danezis2009sybilinfer} is an RW and Bayesian inference-based algorithm for FAD. It performs multiple traces starting from a known real account, storing only the start and endpoints of each trace. In OSNs, based on homophily assumptions, traces that start from known real accounts tend to stay within well-connected real regions, compared to fake accounts, which are less connected to the real region. As a result, a trace starting from a real account is less likely to end in the fake region. \textbf{SybilInfer} estimates the probability that each node belongs to which class (real or fake) using Bayes’ theorem, given a set of traces initiated from known real accounts.

Some RW-based methods provide a ranking for accounts based on their likelihood of being fake, rather than just offering a binary classification. \textbf{SybilRank}~\cite{cao2012aiding} uses RW with short traces, starting from known real accounts and measuring the landing probabilities on other nodes to provide the ranking. \textbf{SybilRank} used the iterative version of RW and then ranks the nodes based on their final scores.

Focusing on locality in graphs has led to significant advancements in FAD. While many methods rely on global structural properties, some research demonstrates that leveraging localised graph characteristics improves FAD.
\textbf{SybilDefender}~\cite{wei2012sybildefender} proposed locality-aware FAD by utilising shorter traces and focusing on local community structures.
Based on the homophily assumption, \textbf{SybilDefender} posits that the limited number of edges between the real and fake regions causes traces from fake accounts to remain trapped within the fake region. Additionally, assuming the fake region is significantly smaller than the real region means a trace originating from a real account will traverse more unique nodes than one from a fake account. The algorithm tracks the number of unique nodes visited multiple times within a given trace. If traces from node $v$ remain trapped in a small region, it is classified as fake; otherwise, it is considered a real account.

Alvisi et al.~\cite{alvisi2013sok} categorised and evaluated the papers in this line of work, highlighting their evolution, strengths, and limitations.

Later, Jia et al.~\cite{jia2017random} proposed \textbf{SybilWalk}, a more advanced RW-based FAD, which simultaneously incorporates both labelled real and fake accounts into the RW framework. Including both label types, particularly in lower homophily, enhances robustness to label noise while preserving scalability.
SybilWalk first augments the graph by introducing two new nodes, referred to as label nodes, denoted as $l_s$ (the fake label node) and $l_b$ (the real label node). These nodes are connected to the network's labelled fake and real accounts.
For each node $ u $, SybilWalk defines the badness score $ p_u $ as the probability that an RW originating from $ u $ reaches $ l_s $ before $ l_b $.
SybilWalk utilises an iterative approach for RW. Initially ($t = 0$), the fake label node score is assigned $ p_{l_s} = 1 $, the real label node is assigned $ p_{l_b} = 0 $, and all other nodes are initialised with $ p_u = 0.5 $. The iterative computation continues until $ p $ converges. Finally, a node $ u $ is classified as fake if $ p_u > 0.5 $ and real otherwise.

\subsection{ Probabilistic Inference and Message Passing Methods}
\label{subsec:cls:prop}

This section focuses on methods that utilise probabilistic models, particularly probabilistic graphical inference, with an emphasis on LBP. These methods iteratively propagate probabilities or scores (the likelihood of nodes being fake accounts) based on local priors and global structure, modelling relationships between nodes probabilistically. This approach enables more robust classification, particularly in cases of weak homophily. By continuously updating beliefs across the network, these methods refine node classifications and enhance the accuracy of FAD. We use the term ``score'' to refer to node-level beliefs in LBP for consistency, as different studies may employ varying terminology (e.g., likelihoods and probabilities).

\paragraph{Markov Random Field (MRF).}
The MRF is a probabilistic graphical model that represents dependencies between variables using a graph structure. In an MRF, each node corresponds to a random variable, and edges capture the probabilistic dependencies between them.
To perform inference over nodes with complex dependencies, these dependencies are simplified by assuming the Markov property, which states that a node is conditionally independent of all other nodes given its neighbors~\cite{el2020similcatch}.
In the undirected graph, the local Markov property is formally expressed as:
$P(X_u \mid X_{V \setminus \{u\}}) = P(X_u \mid X_{\Gamma(u) \setminus \{u\}})$,
meaning that a node’s state depends only on its neighbors~\cite{el2020similcatch}.

Given graph $G = (V, E)$, the joint probability distribution over the variables $X = \{x_v \mid v \in V\}$ can be defined as Equation~\eqref{eq:mrf_distribution}, where $ \phi_v(x_v) $ is the node potential capturing prior score about node $ v $, and $ \psi_{u,v}(x_u, x_v) $ is the edge potential modelling interactions between connected nodes. In addition, $ Z $ is the partition function ensuring normalisation~\cite{gong2014sybilbelief}.

\begin{equation} P(X) = \frac{1}{Z} \prod_{v \in V} \phi_v(x_v) \prod_{(u,v) \in E} \psi_{u,v}(x_u, x_v) \label{eq:mrf_distribution} \end{equation}

\paragraph{Loopy Belief Propagation (LBP).}
LBP is an iterative algorithm for approximate inference in MRFs when the graph contains loops. It passes messages between neighbouring nodes, refining the probabilities of different states over multiple iterations. The message sent from node $ u $ to node $ v $ at iteration $ t $ is based on Equation~\eqref{eq:lbp_message}.
\begin{equation} m_{u,v}^{(t)}(x_v) = \sum_{x_u} \phi_u(x_u) \psi_{u,v}(x_u, x_v) \prod_{w \in \Gamma(u) \setminus v} m_{w,u}^{(t-1)}(x_u),\label{eq:lbp_message} \end{equation}

After the messages stabilise or a stopping condition is met, the final score of node $ v $ is estimated based on Equation~\eqref{eq:marginal_probability}, which provides an approximate probability distribution for each node’s state.

\begin{equation} P(x_v) \propto \phi_v(x_v) \prod_{u \in \Gamma(v)} m_{u,v}(x_v).\label{eq:marginal_probability} \end{equation}

\textbf{SybilBelief}~\cite{gong2014sybilbelief} applies LBP for FAD by formulating the problem as a probabilistic graphical model. The OSN is represented as an undirected graph, where each node is assigned a binary label: real account ($+1$) or fake account ($-1$). Known real and fake accounts act as labelled data. SybilBelief models homophily through edge potentials that favour connected nodes with the same label.
Using LBP iteratively propagates scores from labelled nodes to the rest, refining their probability of being a fake or real account. If only real account labels are available, a boosting strategy is used, where multiple times different random nodes are temporarily treated as fake accounts to improve classification accuracy.

\textbf{SybilFrame}~\cite{gao2015sybilframe} improves upon SybilBelief by addressing its failure in networks with lower homophily. Its novelty lies in a multi-stage classification framework that integrates local information about node and edge priors with global structural inference. The method first extracts local information (personal features like incoming/outgoing request acceptance ratios and graph-based features like clustering coefficient and Jaccard similarity) to compute prior scores, then applies LBP for probabilistic inference. 
% \textcolor{brown}{This paper uses an ML model for computing prior score from its features, but the main body of the SybilFrame is a classical LBP model.}

\textbf{SybilSCAR}~\cite{wang2017sybilscar} unifies RW and LBP methods under a single framework, mitigating the computational complexity of edge-specific message passing. This novel approach enables efficient FAD by iteratively updating each node's posterior probability based on the influences of its neighbours and prior knowledge.
The key innovation of SybilSCAR lies in its local rule, which integrates the advantages of RW-based and LBP-based methods while overcoming their limitations. Traditional RW-based methods struggle to simultaneously incorporate both known fake and real accounts, resulting in lower accuracy and higher sensitivity to label noise. LBP-based methods, although more robust, suffer from scalability issues and lack guaranteed convergence due to their reliance on iterative message passing across edges, which means LBP might oscillate on graphs with loops. SybilSCAR resolves these challenges by introducing a local update rule that models a node’s posterior probability as a function of its neighbours’ influences in a multiplicative yet computationally efficient manner.

\textbf{SybilHP}~\cite{lu2023sybilhp} is an LBP-based method designed for directed OSNs, where it improves detection by introducing adaptive and direction-sensitive edge potentials. Unlike traditional approaches with fixed homophily, where edge strengths are predefined and remain constant across the graph, SybilHP updates edge strengths iteratively based on local label distributions.
Given a directed graph $G = (V, E)$, nodes represent accounts with labels $x_u \in \{-1, 1\}$ for real and fake accounts, respectively. Initial prior scores are set as $p(x_u = 0)$ for known real, $1$ for known fake, and $0.5$ for unlabelled nodes.
At each iteration, SybilHP computes a directional homophily score for node $u$ as Equation~\eqref{eq:homophily_strength_sybilhp}, where $\Gamma_{\text{dir}}(u)$ denotes either the set of in-neighbors or out-neighbors of node $u$ ($\text{dir} \in \{\text{in}, \text{out}\}$). In addition, $p^{(t)}(x_v = l)$ is the estimated probability at iteration $t$ that neighbour $v$ has label $l$. The choice of $l$ depends on the direction: for $\textbf{dir} = \text{in}$, $l = -1$ (real account), and for $\textbf{dir} = \text{out}$, $l = 1$ (fake account). 

\begin{equation} c_u^{(\text{dir}, t)} = \frac{1}{|\Gamma_{\text{dir}}(u)|} \sum_{v \in \Gamma_{\text{dir}}(u)} p^{(t)}(x_v = l) \label{eq:homophily_strength_sybilhp} \end{equation}

\subsection{Structural Preprocessing and Feature Integration Method}
\label{subsec:cls:hyb} 

While classical approaches primarily rely on the graph structure or a small set of labelled data, newer methods aim to enhance detection by either modifying the graph or incorporating additional features. These improvements help make the network structure more refined and informative. Examples include removing noisy edges or adding more features such as account activity and interactions.
% This makes FAD more accurate and better suited to real-world scenarios.

Some methods preprocess the graph to remove weak edges or integrate additional interaction information, making the structure more suitable for FAD and enhancing score propagation. Effendy and Yap~\cite{effendy2017strong} aimed to increase homophily in OSNs by reinforcing connections within real account clusters. This paper introduces the Strong Link Graph (\textbf{SLG}), which removes weak edges while preserving strong edges. This transformation reduces the influence of attack edges (attack edges are those from fake accounts to real accounts), which are generally weaker based on the homophily assumption, while maintaining connections between real accounts, making FAD more effective.

\textbf{SybilHunter}~\cite{mao2022sybilhunter} improved SybilRank by refining score initialisation and incorporating both network structure and personal features. Unlike SybilRank, which assigns uniform initial scores to known nodes, SybilHunter derives them from extracted features, such as the accepted request ratio and follower-following ratios.
SybilHunter first extracted features and utilised them to estimate credibility. Node credibility represents the probability that an account is real, while edge credibility reflects the likelihood that both endpoints of an edge have the same label.
SybilHunter builds a weighted graph using edge credibility, removes weak edges similar to the SLG~\cite{effendy2017strong} method, which was described above, and applies a weighted RW. The initial scores are based on node credibility, and the final scores determine the ranking of fake accounts.

Unlike previous methods that rely solely on the friendships and the homophily assumption, \textbf{Sybil\_SAN}~\cite{zhang2022enhancing} improves ranking-based FAD by preprocessing the network into a Social-Activity Network (SAN), integrating both friendships and interactions (e.g., mention, retweet, and reply), which enhances score propagation and reduces reliance on the limited-attack-edge assumption. Sybil\_SAN leverages RWs on the SAN to perform more accurate detection by exploring the network structure 
while reducing the impact of attack edges by propagating scores primarily through interaction paths where real accounts are less likely to engage with fake accounts,
making it more robust in lower homophily networks where fake accounts form many connections with real accounts.

Jethava and Rao~\cite{jethava2022user} utilised a combination of important relationships between accounts, including common groups, mutual likes, and betweenness centrality, to identify probable fake accounts. Their approach focuses on detecting attack edges and fake accounts by measuring relationship strength, emphasising the significance of this measurement in FAD. Their method first evaluates account interaction patterns using similarity metrics to compute relationship strength. Then, it leverages betweenness-centrality to distinguish fake accounts from real ones, identifying low-strength connections as attack edges and isolating fake accounts.

Dehkordi and Zehmakan~\cite{dehkordi2025efficient} introduced the notion of account resistance, which refers to the probability that real accounts reject incoming requests from fake accounts. They proposed algorithms that leverage resistance information to identify more real accounts and detect potential attack edges. These resistance-aware strategies can serve as preprocessing steps to increase the number of known real nodes and, by decreasing the weights of potential attack edges, improve graph homophily, ultimately enhancing the effectiveness of existing detection methods, such as SybilSCAR.
% \ali{Dear @Ahad, should we say \textbf{WE} here or just leave it like this?} \anz{Both styles are OK, but they is more common.} 
%\ali{ok, Thank you.}

In contrast to Sybil\_SAN, which focuses on sanitising attack edges before applying detection, \textbf{SybilEdge}~\cite{breuer2020sybiledge} adapts dynamically by analysing 
personal features (how fake and real accounts choose their friend request targets and how these targets respond in real-time). SybilEdge does not explicitly assign weights to edges in the traditional sense but implicitly assigns them based on the probability that a request sender’s choice of targets resembles that of a fake or real account. After assigning these weights, SybilEdge computes each account's likelihood of being fake by aggregating target selection and response behaviours, which enables early detection, identifying fake accounts even when they have made only a few connections before they can blend into the network.

Talking about early detection, \textbf{PreAttack}~\cite{breuer2023preattack} introduces a probabilistic framework that models how new accounts initiate social connections using a multi-class Preferential Attachment model. It monitors the friend request behaviour of new accounts and identifies fake ones by detecting significant differences from the statistical patterns expected of real accounts. This preemptive strategy enables early detection, preventing fake accounts from gaining network influence.

Unlike previous methods that rely on fixed label sets and are vulnerable to evolving manipulation strategies, \textbf{RICC}~\cite{shin2023ricc} enhances FAD by leveraging collective classification. In this method, a node's label is influenced by its features and the labels of its neighbours. RICC iteratively refines scores that represent the likelihood of accounts being real. It stabilises classification results by incorporating multiple rounds of randomly sampled subsets of prior scores. In each round, it minimises the difference between posterior scores computed from the graph and those derived from a randomly sampled subset. This iterative adjustment enhances robustness against changes in fake account strategies and ensures stability across diverse graph structures.

Talking about labels, while many FAD methods heavily rely on labelled data, \textbf{SybilBlind}~\cite{wang2018sybilblind} eliminates this dependency, making it more adaptable to real-world networks where ground-truth labels are scarce. Instead of relying on pre-annotated fake or real accounts, SybilBlind generates pseudo-labels using random sampling and aggregates multiple detection trials to improve accuracy.
SybilBlind begins by randomly selecting two subsets of nodes and assigning pseudo-labels, treating one subset as real nodes and the other as fake nodes. An FAD algorithm, such as SybilSCAR, is then applied to these new pseudo-labelled datasets.

While SybilSCAR and SybilFuse combined elements of RW and LBP, Furutani et al.~\cite{furutani2020sybil} introduced a Graph Signal Processing (GSP) perspective, interpreting FAD as a low-pass filtering task. In this view, effective detection preserves low-frequency components reflecting community structures while suppressing high-frequency noise, particularly from high-degree nodes. This explains the bias of RW methods and the success of LBP in well-clustered graphs.

Building on this, \textbf{SybilHeat}~\cite{furutani2023interpreting} unified various FAD methods (e.g., SybilRank, SybilWalk, SybilSCAR, SybilBelief) under a GSP framework. By analysing the spectral properties of graph matrices such as the adjacency and laplacian matrices, SybilHeat applies an optimised filter that improves community detection and reduces noise across diverse network conditions.

% Furutani et al.~\cite{furutani2023interpreting} introduced \textbf{SybilHeat}, which unifies multiple existing approaches, such as SybilRank, SybilWalk, SybilSCAR, and SybilBelief, under a common mathematical framework. The matrices used in these models, such as the adjacency and Laplacian matrices, define how signals spread across a graph. Their eigenvalues and eigenvectors determine how effectively low-frequency components reflect community structure and can be separated from high-frequency noise. By studying different filtering techniques and analysing these matrices, Furutani et al. gained new insights into why some FAD methods work well in specific networks but fail in others. SybilHeat improves this process by applying an optimised filter that enhances community detection while reducing noise.

Despite advancements in ML and DL, some researchers continue to explore the classical algorithms, emphasising the advantages of a graph-theoretic perspective while integrating behavioural cues to adapt to evolving fake account strategies. Each generation of methods has addressed new challenges, from reducing computational overhead to enhancing resilience against label noise and adaptive attackers. A summary of classic methods and their key contributions is provided in Table~\ref{tab:classic_paper_summary}.

\begin{table*}[ht!]
    \centering
    \caption{Summary of classical methods (Section~\ref{sec:classic_algos}) with their main idea.}
    \label{tab:classic_paper_summary}
    \renewcommand{\arraystretch}{1} 
    \begin{tabularx}{\textwidth}{|>{\hsize=0.18\hsize}X| >{\hsize=0.27\hsize}X|>{\hsize=0.55\hsize}X|}  
        \toprule
        \textbf{Paper} & \textbf{Primary Technique} & \textbf{Main Idea} \\
        \midrule
        Yang et al.~\cite{yang2014uncovering} & Empirical Study + Bayesian Inference & Challenging homophily assumption and proposing Bayesian FAD. \\ \hline
        SybilGuard~\cite{yu2006sybilguard} & RW & Introducing RW to separate fake and real accounts using the homophily assumption. \\ \hline
        SybilLimit~\cite{yu2008sybillimit} & RW & Improving scalability by reducing the number of required RWs. \\ \hline
        SybilInfer~\cite{danezis2009sybilinfer} & RW + Bayesian Inference & Estimating node probabilities using Bayesian inference based on RW traces. \\ \hline
        SybilRank~\cite{cao2012aiding} & RW-based Ranking & Providing a ranking system for prioritising manual inspections of fake accounts. \\ \hline
        SybilDefender~\cite{wei2012sybildefender} & RW & Utilising short RWs to utilise local community structures effectively. \\ \hline
        SybilWalk~\cite{jia2017random} & RW & Incorporating both labelled real and fake nodes into RWs to enhance the robustness. \\ \hline
        SybilBelief~\cite{gong2014sybilbelief} & LBP & Applying belief propagation to refine node classification probabilities iteratively. \\ \hline
        SybilFrame~\cite{gao2015sybilframe} & Weighted LBP  & Integrating personal and graph-based features to have weighted LBP and running the multi-stage classification. \\ \hline
        SybilSCAR~\cite{wang2017sybilscar} & RW + LBP Hybrid & Unifying RW and belief propagation techniques to enhance computational efficiency. \\ \hline
        SybilHP~\cite{lu2023sybilhp} & Adaptive Directed LBP & Adapting belief propagation dynamically by changing edge potentials in LBP. \\ \hline
        Effendy et al.~\cite{effendy2017strong} & Graph Refinement & Filtering weak edges to increase homophily in the graph. \\ \hline
        SybilHunter~\cite{mao2022sybilhunter} & Weighted RW-based Ranking & Refining RW initialisation with feature-based scores. \\ \hline
        Sybil\_SAN~\cite{zhang2022enhancing} & Activity-based RW & Enhancing score propagation in the graph by integrating social interactions alongside friendships for RW.\\ \hline
        Jethava et al.~\cite{jethava2022user} & Graph Feature Analysis  & Identifying fake accounts using betweenness centrality and interaction metrics. \\ \hline
        Dehkordi and Zehmakan~\cite{dehkordi2025efficient} & Graph Preprocessing & Introducing the notion of account resistance and preprocessing based on that. \\ \hline
        SybilEdge~\cite{breuer2020sybiledge} & Personal Feature-based Early Detection & Early detection by analysing friend request patterns and responses before they establish many connections. \\ \hline
        PreAttack~\cite{breuer2023preattack} & Modelling Account Behaviours For Early Detection. & Early detection by modelling how new accounts send and receive friend requests using a multi-class Preferential Attachment framework.\\ \hline
        RICC~\cite{shin2023ricc} & Collective Classification & Refining classification iteratively with random sampling. \\ \hline
        SybilBlind~\cite{wang2018sybilblind} & Pseudo-labelling & Detecting without labelled data by generating pseudo-labels iteratively. \\ \hline
        Furutani et al.~\cite{furutani2020sybil} & Graph Signal Processing & Interpreting RW and belief propagation methods through low-pass filtering. \\ \hline
        SybilHeat~\cite{furutani2023interpreting} & Graph Signal Processing & Unifying multiple detection methods using optimised graph signal filtering. \\ 
        \bottomrule
    \end{tabularx}
\end{table*}

\section{Traditional Machine Learning} 
\label{sec:ML}

This section is focused on traditional ML methods for FAD, with DL methods excluded. These approaches are characterised by their simplicity, interpretability, and efficiency. Key contributions from the literature are reviewed, with an emphasis on recent works.

Some prior research in FAD has focused on extracting discriminative features from profiles and graph-based features. These methods leverage classification algorithms such as SVM, DT, RF, and logistic regression. For instance, \textbf{TSD}~\cite{alsaleh2014tsd} extracts profile and content features, proving that even basic ML models can detect fake accounts when given the right features. Later, Erşahin et al.~\cite{ercsahin2017twitter} demonstrated that applying a Naïve Bayes model with discretised numerical features improves detection accuracy. Regarding the use of more complex features,
BotShape~\cite{wu2023botshape} enhanced detection accuracy by analysing posting activity over time, demonstrating that time-based posting patterns reveal fake accounts more effectively than static profile features alone.
Furthermore, Sallah et al.~\cite{sallah2024efficient} proposed a feature selection method using the Non-Dominated Sorting Genetic Algorithm II (\textbf{NSGA-II}). This multi-objective optimisation algorithm simultaneously maximises classification accuracy and reduces the number of selected features, effectively balancing model performance and complexity.

The increasing sophistication of fake accounts has driven researchers to focus on exploiting network structures for detection. Graph-based approaches leverage connectivity patterns to uncover coordinated activity among fake accounts, providing a more resilient and scalable method.
These approaches range from simple graph similarity measures to advanced higher-order representations and probabilistic methods. 
To organise the literature, we categorise existing research into two main areas: graph-based ML methods, which focus on structural features, connectivity patterns, and local or global graph-based learning strategies; and higher-order graph techniques and probabilistic methods, which incorporate advanced representations such as multi-relational graphs, belief propagation, and structural entropy to enhance detection accuracy and address the limitations of simpler methods.

\subsection{Graph-Based ML Methods}
\label{subsec:ML-Graph}

A straightforward approach in graph-based methods for FAD is utilising graph-based features or basic structural analysis techniques.
Focusing on feature-based methods, Asghari et al.~\cite{asghari2022using} analysed various graph-based features, specifically node-level measures such as degree, eigenvector centrality, harmonic centrality, and local reaching centrality, as well as their correlations. Their study demonstrated that these structural properties can effectively differentiate real and fake accounts.
Mohammadrezaei et al.~\cite{mohammadrezaei2018identifying} proposed a graph-based detection framework that utilises similarity measures, including the Jaccard index, cosine similarity, and common friends, in conjunction with Principal Component Analysis (PCA) for dimensionality reduction. Their results demonstrated that analysing account connectivity through these similarity measures significantly improves classification accuracy, particularly in identifying fake accounts.

Beyond simple connectivity analysis, researchers have explored ego networks for nodes (subgraphs centred around nodes, including their direct connections and the relationships among them) and weak link structures to detect fake accounts based on their local connectivity. Bebensee et al.~\cite{bebensee2021leveraging} demonstrated that fake accounts often form densely connected communities while maintaining only a few weak edges to real accounts (similar to the homophily assumption), making them structurally distinct in social graphs. Their method creates an ego network and then extracts graph-based features such as density (the proportion of existing edges among neighbours) and reciprocity (the likelihood of mutual connections) to distinguish real from fake accounts. 

Another representative method that combines structural analysis with ML is SybilFUSE~\cite{gao2018sybilfuse}. This approach utilises local classifiers, such as SVM or logistic regression models, trained on node- and edge-level features, including incoming and outgoing friend request acceptance ratios and local clustering coefficients. These classifiers produce scores that reflect the likelihood of a node being real or an edge not from a fake account.
In contrast to traditional FAD that rely solely on RWs or LBP, SybilFUSE integrates these global propagation techniques with the outputs of local ML models. In the weighted RW setting, local scores are propagated through the network while reducing the influence of suspected attack edges. In the weighted LBP variant, the network is treated as a Markov Random Field, and belief scores are refined using edge potentials derived from the scores assigned by local classifiers.
By combining local classification with global, structure-aware propagation, SybilFUSE relaxes several strong assumptions commonly made in prior work, such as strict homophily or a limited number of attack edges. This hybrid design improves robustness to label noise and maintains high performance even in low-homophily or structurally complex networks.

Given the limited availability of labelled nodes, semi-supervised and unsupervised methods are crucial for effective FAD.
BalaAnand et al.~\cite{balaanand2019enhanced} proposed an Enhanced Graph-based Semi-Supervised Learning Algorithm (\textbf{EGSLA}) to detect fake accounts on Twitter. Instead of relying purely on supervised methods like k-NN or SVM, their method improves performance by leveraging network structure and relationships among accounts in a social graph. The proposed method employs a graph-based semi-supervised learning algorithm that utilises the graph Laplacian for classification and decision-making, incorporating label propagation to distinguish fake accounts. It spreads known labels to unlabelled nodes based on connectivity, enabling effective classification with minimal labelled data.

% Kagan et al.~\cite{kagan2018generic} introduced an unsupervised anomaly detection method using link prediction. Their two-layered meta-classifier identifies anomalous nodes, including fake accounts, by detecting unlikely network connections based on typical graph structures. First, an RF classifier is trained for link prediction, utilising features such as Jaccard’s coefficient and the Adamic-Adar index to estimate the probability of edges existing between nodes. Connections with low probabilities are considered improbable, indicating unusual or unnatural relationships. Then, an anomaly detection model is applied to features derived from these probabilities to classify nodes as anomalies.

While most researchers have focused on after extended activity detection, proactive strategies aim to detect fake accounts at registration time detection or through early detection.
\textbf{Ianus}~\cite{yuan2019ianus} is a method for FAD at registration time, leveraging personal features, including shared features between different accounts (e.g., shared IP, phone numbers, and devices) and anomalous features (e.g., late-night sign-ups, rare OS versions, and geolocation inconsistencies). It constructs a weighted registration graph, where nodes represent accounts and edges are weighted by the scores, which quantify the likelihood that two accounts belong to the same fake accounts cluster. These scores are computed using a logistic regression model trained on the mentioned features. Based on Ianus, fake accounts form densely connected subgraphs due to shared registration attributes, while real accounts remain sparsely linked. Ianus performs FAD using two methods: First, community detection (the Louvain method) identifies tightly connected clusters of fake accounts. Second, the weighted node degree approach estimates the strength of an account's connections to others by computing the sum of its connection scores.

\subsection{Higher-Order Graph Techniques and Probabilistic Methods} 
\label{subsec:ML-HigherOrder}

To overcome the limitations of simple methods, researchers have explored higher-order graph techniques and utilised probabilistic methods in combination with ML models. These approaches incorporate community structure, belief propagation, entropy-based reasoning, and multi-relational patterns to improve ML-based methods. They often combine unsupervised methods to handle the label limitation problem.

Liang et al.~\cite{liang2021unveiling} proposed \textbf{UFA} (Unveiling Fake Accounts), an unsupervised ML system for at registration time detection. UFA constructs a registration-feature bigraph, where nodes represent accounts, and their personal features include registration attributes (e.g., IP addresses, device fingerprints). It then applies unsupervised weight learning, initialising anomaly scores for accounts and features based on statistical deviations from normal registration patterns. These initial weights are refined through linearised belief propagation, a graph-based iterative message passing algorithm that propagates anomaly scores across the bigraph. Next, UFA transforms the bigraph into a registration graph, where nodes represent accounts and edges are created between accounts with high registration similarity. Fake accounts tend to form densely connected communities, while real accounts remain more sparsely connected. UFA employs community detection (the Louvain method) to detect fake accounts, which identifies clusters of densely connected fake accounts. By leveraging graph-based learning techniques, combining belief propagation and community detection, UFA detects fake accounts at registration time before they engage in fake account activities.

Regarding the time of detection, for the early detection theme, Zhang et al.~\cite{zhang2023gufad} proposed \textbf{GUFAD}, a framework for FAD at both the registration and login stages. It leverages graph analysis, community detection, graph embedding, and clustering techniques to identify fake accounts without requiring labelled data.
The framework constructs an account-feature graph, applies community detection (using the Louvain algorithm) to group potential fake accounts, and utilises \textbf{Graph2Vec}~\cite{narayanan2017graph2vec} to learn representations of the detected communities. It then applies K-Means for clustering to classify them as fake or real. To enhance detection, GUFAD incorporates feature engineering with personal features, such as IP addresses, passwords and dynamic features that capture patterns over time. The time-based sliding window extracts features from account activity within fixed intervals, detecting sudden increases in fake activity. Features such as unique IPs and shared passwords exhibiting synchronised behaviour. The count-based sliding window segments data based on a fixed number of records, helping to identify unusually large batches of fake accounts appearing within a short time frame.

Transitioning from account-based clustering to structure-aware reasoning, \textbf{SimilCatch}~\cite{el2020similcatch} enhances FAD by combining supervised learning with probabilistic graphical models to detect spammers on Twitter. Unlike approaches that rely on social connections, it builds a bipartite graph of shared content to model the similarities between accounts involved in coordinated spam campaigns. Initial account probabilities are generated using classifiers (SVM, RF, or logistic regression), which are prior beliefs in an MRF. The LBP then refines these probabilities by propagating information across linked nodes. To measure structural uncertainty, SimilCatch uses structural entropy based on Equation~\eqref{eq:struc-ent}, where $p_i$ denotes the probability mass of structural component $i$.

\begin{equation}
    H(G) = - \sum_{i} p_i \log p_i
    \label{eq:struc-ent}
\end{equation}

Building on entropy-based detection, \textbf{UnDBot}~\cite{peng2024unsupervised} proposes an unsupervised approach using a multi-relational graph that encodes different features (e.g., posting frequency and follow ratios). It applies Hierarchical Structural Entropy (HSE), which recursively partitions the graph into subgraphs while minimising entropy, revealing irregular structures associated with fake accounts. The HSE is defined as Equation~\eqref{eq:H-struc-ent}, where $\mathcal{P}$ is a partition of the graph, $p_C$ is the probability of subgraph $C$ which is $p_C = \frac{\text{vol}(C)}{\text{vol}(G)}$ where graph volume is $\text{vol}(G) = \sum_{v \in V} \deg(v)$, and $H(C)$ is its entropy.

Unlike relying on static features, UnDBot emphasises long-term behavioural consistency by tracking account interactions across contexts, improving robustness against evasion tactics.
\begin{equation}
    H(G) = \sum_{C \in \mathcal{P}} p_C H(C) - \sum_{C \in \mathcal{P}} p_C \log p_C
    \label{eq:H-struc-ent}
\end{equation}

% UnDBot shifts from static graph-based features to long-term analysis to counter the evolution of fake account strategies. Fake accounts adapt their interactions to evade detection by mimicking human-like activity. Instead of relying solely on direct connections, UnDBot models multiple mentioned features to capture the patterns that persist over time. By constructing a transition model that tracks account interactions across different relational contexts, UnDBot detects persistent anomalies, ensuring resilience against evolving evasion strategies.

Expanding on this, \textbf{SIASM}~\cite{zeng2024adversarial} introduces a proactive framework that models how fake accounts adapt their behaviour to avoid detection. It models the OSN as a multi-relational graph, then uses RL to optimise social activity and convert it into a graph with a single relation type. Structural entropy is minimised throughout this transformation to preserve meaningful structure. SIASM uses first-order structural entropy ($H_1$) based on node degrees as Equation~\eqref{eq:ent_siasm}, which is a special case of Equation~\eqref{eq:struc-ent}.

\begin{equation}
    H_1(G) = -\sum_{v \in V} \frac{\deg(v)}{\text{vol}(G)} \log_2 \frac{\deg(v)}{\text{vol}(G)}
    \label{eq:ent_siasm}
\end{equation}

Nguyen et al.~\cite{nguyen2020bot} introduced an FAD using persistent homology, a graph-based feature (e.g., connected components, loops, and holes) analysis tool that captures complex patterns in networks. Instead of using simple graph-based features like node degree, it builds ego networks for each account, representing their direct connections. These networks are then converted into simplicial complexes (generalised graph structures that include not just edges but also higher-dimensional connections), which help analyse loops and missing connections. The method tracks how these patterns appear and disappear over different connection strengths, encoding them into embeddings. Then, the distance measure compares these embeddings, and finally, the $k$-nearest neighbours classifier is used to distinguish between real and fake accounts.

\textbf{BotFinder}~\cite{li2023botfinder} integrates graph embedding methods, such as Node2Vec~\cite{grover2016node2vec}, and community detection (using the Louvain method) to analyse structural relationships between accounts and identify similarities. The study combines supervised learning (feature engineering and classification) with unsupervised community detection, where graph-based label diffusion refines initial predictions. This hybrid approach significantly improves FAD by leveraging individual features and network structure to enhance classification accuracy.

In addition, ML-based methods have been explored to analyse how fake accounts manipulate social structures. Zhou et al.~\cite{zhou2020efficient} introduced a supervised learning approach that predicts victim accounts using a classifier trained on extracted data. Their method assigns a victim score to each account and, based on the predictions, modifies edge weights in the social graph. This refinement enhances a subsequent score propagation algorithm, effectively filtering out fake accounts. 

\textbf{Íntegro}~\cite{boshmaf2015integro} identifies victims using an ML classifier trained on features like friend count, messages sent, and account age. Based on labelled training data, an RF model distinguishes victims (accounts that unknowingly befriend fake accounts) from others. Each account receives a probability score, and those above a threshold are labelled potential victims. The graph is then reweighted by lowering the edge weights of predicted victims, making it easier to distinguish between fake and real accounts. This approach improves detection robustness, as attackers cannot control victim behaviour.

Graph-based ML methods for FAD have evolved from simple feature analysis to sophisticated methods incorporating embeddings, probabilistic reasoning, and higher-order graph representations. Integrating community-aware and entropy-based techniques has further enhanced model scalability and interoperability. 
A summary of key graph-based ML models and their technical contributions is provided in Table~\ref{tab:graph_ml_summary}.

\begin{table*}[ht!]
    \centering
    \caption{Summary of traditional ML-based methods (Section~\ref{sec:ML}) with their main idea.}
    \label{tab:graph_ml_summary}
    \begin{tabularx}{\textwidth}{|>{\hsize=0.16\hsize}X |>{\hsize=0.3\hsize}X |>{\hsize=0.54\hsize}X|}
        \toprule
        \textbf{Paper} & \textbf{Primary Technique} & \textbf{Main Idea} \\
        \midrule
        Asghari et al.~\cite{asghari2022using} & Graph-Based Feature Correlation Analysis & Leveraging node-level measures and correlations for FAD. \\ \hline
        Mohammadrezaei et al.~\cite{mohammadrezaei2018identifying} & Graph Similarity Analysis + PCA & Enhancing classification accuracy with graph-based similarity and dimensionality reduction. \\ \hline
        Bebensee et al.~\cite{bebensee2021leveraging} & Ego Network Analysis + Weak Link Detection & Exploiting ego-network topology differences between real and fake accounts. \\ \hline
        SybilFuse~\cite{gao2018sybilfuse} & SVM + Weighted RW + LBP & Combining SVM-based local classifier with weighted RWs and belief propagation.\\ \hline
        EGSLA~\cite{balaanand2019enhanced} & Enhanced Semi-Supervised Learning, Graph Laplacian & Leveraging labelled and unlabelled data with graph structure for FAD. \\ \hline
        Ianus~\cite{yuan2019ianus} & Registration-Based Weighted Graph + Community Detection & Detecting fake accounts at registration time using shared attributes (IP, device) and abnormal behavioural patterns. \\ \hline
        UFA~\cite{liang2021unveiling} & Registration-Feature Bigraph + Belief Propagation & Detecting fake accounts by identifying outlier registration patterns without labelled data. \\ \hline
        GUFAD~\cite{zhang2023gufad} & Graph Embedding + Clustering + Multi-Granularity Analysis & Capturing dynamic fake behaviour via feature coupling and time-based aggregation. \\ \hline
        SimilCatch~\cite{el2020similcatch} & Markov Random Fields + Bipartite Content-account Graph & Propagating beliefs to detect social spammers based on content similarity. \\ \hline
        UnDBot~\cite{peng2024unsupervised} & Structural Entropy + Unsupervised Learning & Detecting social fake accounts using entropy-based structural information. \\ \hline
        SIASM~\cite{zeng2024adversarial} & Structural Entropy + Adversarial Modelling & Simulating adversarial fake account behaviours using hierarchical community structure. \\ \hline
        Nguyen et al.~\cite{nguyen2020bot} & Persistent Homology + Ego Networks & Leveraging topological features of ego networks for FAD. \\  \hline 
        BotFinder~\cite{li2023botfinder} & Graph Embedding (Node2Vec), Community Detection & Combining feature engineering, embedding, and label propagation for FAD. \\ \hline
        Zhou et al.~\cite{zhou2020efficient} & Victim Prediction + Score Propagation & Improving FAD by reweighting graph edges based on victim classification. \\ \hline
        Íntegro~\cite{boshmaf2015integro} & Hybrid Feature-Graph method + Supervised Ranking & Integrating victim prediction into graph-based ranking for robust FAD. \\ 
        \bottomrule
    \end{tabularx}
\end{table*}

\section{Deep Learning} 
\label{sec:DL}

DL has emerged as a powerful approach for FAD in OSNs. It effectively addresses the limitations of traditional ML models that rely on handcrafted features and struggle to adapt to evolving fake account attack strategies. In contrast, DL models can automatically extract complex patterns from raw data, thereby improving detection accuracy and reducing the need for manual intervention.  
Graph-based DL methods, particularly GNNs, have gained significant attention in FAD due to their ability to capture relational and structural dependencies in OSNs, making them particularly well-suited for FAD. Various GNN architectures, such as Graph Convolutional Network (GCN), Graph Attention Network (GAT), and GraphSAGE, have been extensively explored for FAD, demonstrating improved performance in identifying fake accounts.

Several studies have demonstrated the effectiveness of using GNN variants in FAD. For instance, Yang and Zheng~\cite{yang2020fake} proposed an attention-based GCN that captures aggregation patterns in OSNs while reweighting neighbour importance using an attention mechanism. Heeb et al.\cite{heeb2024sybil} introduced \textbf{SYBILGAT}, a GAT-based method that leverages attention to highlight relevant graph-based features and supports inductive learning by pretraining on sampled subgraphs. Furthermore, Khan et al.~\cite{khan2024graph} proposed a GraphSAGE-based framework that incorporates profile, content, and ego-network features, showing the potential of inductive structural embeddings for detecting fake accounts in unstructured environments. In addition, several studies proposed methods to enhance the robustness of GNN-based models. For instance, \textbf{OS-GNN}~\cite{shi2024over} is a method that addresses severe class imbalance by creating synthesised minority-class (fake) samples in feature space.

While message passing GNNs have significantly advanced FAD by capturing relational and structural dependencies within OSNs, they remain limited in handling label imbalances, adversarial manipulation, heterophilic interactions, multimodal relations, and complex behavioural patterns. To overcome these limitations, recent studies have proposed a diverse set of DL-based methods that enhance the performance of FAD.

\subsection{Deep Learning Extensions of Classical Methods}
\label{subsec:DL:classic}

Several FAD methods enhance classical approaches, such as RW-based methods~\cite{lingam2020social,sun2020trustgcn} and ensemble learning methods (e.g., RF~\cite{shi2023rf} and XGBoost~\cite{tang2024uncovering}). These hybrid methods leverage the strengths of classical techniques while benefiting from the power of DL models.

Some score-based methods integrate graph structure with RW strategies to help identify parts of the network that are more likely to contain real accounts (recall that a score is the likelihood of a node being a fake account).
\textbf{TrustGCN}~\cite{sun2020trustgcn} follows this idea by computing scores from short traces (recall that a trace is a single instance of an RW process) started from known real accounts, assigning higher values to nodes more likely to be real. These scores are then used to weight edges during graph convolution, enabling the GCN to down-weight the influence of likely fake accounts during feature aggregation.

Lingam et al.~\cite{lingam2020social} employed deep autoencoders and an RW model to isolate fake account communities. This approach constructs a weighted signed network graph, where edge weights reflect behavioural similarity and scores. Behavioural similarity is measured along four key aspects: tweet-content similarity, shared URL similarity, interest similarity, and social interaction similarity. Scores are estimated using an RW model, where each node’s score is updated iteratively by aggregating the score of its neighbours. The resulting graph is then embedded using a deep autoencoder framework, which reconstructs community structures more accurately. 

Another approach is to combine classical ensemble techniques, such as RF~\cite{shi2023rf,sanchez2024early}, stacking~\cite{shi2024sstackgnn}, and XGBoost~\cite{tang2024uncovering}, with GNNs to improve FAD methods.
Shi et al.~\cite{shi2023rf} proposed \textbf{RF-GNN}, a method that combines RF ensemble learning with GNNs for FAD. In this approach, RF plays a role by diversifying the training process, much like it does in classical ensemble learning methods. Instead of decision trees, multiple GNN classifiers serve as base learners, each trained on a different subgraph generated through node sampling, feature selection, and edge dropout.
Additionally, a Multi-Layer Perceptron (MLP) is trained on the remaining features not used in the GNN classifiers, ensuring that discarded feature information is still utilised. The outputs of the MLP and GNN classifiers are then aligned. Finally, RF-GNN follows the bagging principle of RF, where predictions from all classifiers are aggregated to make the final decision. This reduces overfitting, increases robustness, and effectively leverages both graph structure and node attributes. It is worth noting that RF-GNN is compatible with various GNN architectures.

S{\'a}nchez-Corcuera et al.~\cite{sanchez2024early} proposed an early detection method using a dynamic graph that continuously updates account interactions to predict future behaviour. Instead of relying on static datasets, their approach identifies potential threats by tracking changes in account embeddings over time. The method employs JODIE, a temporal graph neural network, to forecast account embeddings over time, capturing behavioural patterns before harmful actions occur. These predicted an RF classifier then processes embeddings to distinguish between real and fake accounts.

The next approach to improve account classification performance is to leverage data augmentation to generate diverse training samples. \textbf{SStackGNN}~\cite{shi2024sstackgnn} builds on ensemble learning by employing stacking, a technique where multiple base classifiers are trained independently, and their predictions are then combined by a secondary classifier to produce the final output. In traditional stacking, $K$-fold cross-validation is employed, a method that partitions the training data into $K$ parts and alternates between training and validation models. While this helps mitigate overfitting, it significantly increases computational cost. SStackGNN addresses this challenge by replacing K-fold cross-validation with graph data augmentation, including node dropping, edge perturbation, and feature mixing (combining features from multiple nodes). These augmentations generate diverse graph views, allowing the model to train base classifiers on varied versions of the data.

Tang et al.~\cite{tang2024uncovering} proposed an XGBoost-enhanced GCN approach for FAD. This method first applies XGBoost, a fast and accurate decision tree-based method that ranks features by importance, to select the most informative profile features and reduce redundancy and dimensionality (i.e., the number of features). Selected features are fed into a GCN, integrating them with the network’s structural information through message passing to generate high-level account representations.

Probabilistic frameworks have also been integrated with GCNs to model uncertainties and label dependencies in FAD.
Unlike methods that rely solely on direct neighbourhood aggregation, \textbf{MDGCN}~\cite{deng2022markov} integrates GCN with Adaptive Reward Markov Random Fields (ARMRF) and Conditional Random Fields (CRF) to enhance spammer detection in OSNs. 
Spammers are fake accounts that spread unwanted or deceptive content to manipulate or disrupt social networks.
ARMRF refines predictions by assigning different learnable weights to edges and adjusting the influence of various types of connections (bidirectional, incoming, and outgoing) on an account’s label. Meanwhile, CRF models label dependencies using edges, ensuring consistency between connected accounts by constructing a joint probability distribution over all labels.
MDGCN improves label prediction through an Expectation-Maximisation algorithm, an iterative approach for optimising models with latent variables (the unknown labels of accounts). The expectation step estimates these labels by leveraging the ARMRF layer, refining predictions based on learned dependencies. The maximisation step then optimises a pseudo-likelihood function to update model parameters, improving label consistency over iterations. By combining GCN-based feature extraction with probabilistic modelling of relationships, MDGCN achieves superior spammer detection in highly imbalanced OSNs.

\begin{table*}[ht!]
    \centering
    \caption{DL extensions of classical models (Section~\ref{subsec:DL:classic}). Backbone models are base DL architectures. Features abbreviation: c, g, p, per (content, graph, profile, personal); nc, t (numerical/categorical, textual).}
    % \label{tab:graph_ml_summary}
    %\renewcommand{\arraystretch}{1.2}
    \begin{tabularx}{\textwidth}{|>{\hsize=0.15\hsize}X |>{\hsize=0.12\hsize}X |>{\hsize=0.14\hsize}X |>{\hsize=0.59\hsize}X|}
        \toprule
        \textbf{Paper} & \textbf{Features} & \textbf{Backbone} & \textbf{Main Idea} \\
        \midrule
               
Lingam et al.~\cite{lingam2020social} & p,c/nc,t & Deep Autoencoder & Modelling behavioural similarity and score in a signed Twitter graph and reconstructing communities via deep autoencoders. \\
\hline
TrustGCN~\cite{sun2020trustgcn} & g, per/nc & GCN  & Weighting GCN edges with scores from short RWs on signed graphs. \\
\hline
RF-GNN~\cite{shi2023rf} & p,c/nc,t &  GNN & Aggregating predictions from multiple GNNs trained on randomised subgraphs and features. \\
\hline
Tang et al.~\cite{tang2024uncovering} & p,per/nc & GCN & Integrating top-ranked features from XGBoost with GCN representations.\\
\hline 
S{\'a}nchez-Corcuera et al.~\cite{sanchez2024early} & c/t & RNN & Forecasting accounts behaviour by dynamically modelling their interactions with embedding trajectory prediction. \\
\hline
SStackGNN\cite{shi2024sstackgnn} & p,c/nc,t & GNN & Stacking GNNs trained on graph-augmented views and merging them with a secondary MLP. \\
\hline
MDGCN~\cite{deng2022markov} & g,c/t,nc & GCN & Modelling account relations using graph convolution and a Markov random field. \\
\bottomrule
\end{tabularx}
\end{table*}

\subsection{Heterophily-Aware Methods}
\label{subsec:DL:heterophilic}

Many studies rely on the homophily assumption (nodes with the same label, such as real–real or fake–fake, are more likely to connect). This assumption is embedded in many GNN models, where node features are aggregated from their neighbours, reinforcing local similarities. Meanwhile, connections between fake and real accounts, known as heterophilic edges, degrade the performance of these graph-based methods because message passing on such edges causes feature mixing between fake and real accounts, resulting in blurred class boundaries and more false negatives~\cite{wu2023botscl}.
While some methods explicitly handle heterophily (Section~\ref{subsubsec:edge-type-models}), others implicitly handle heterophily by mitigating its effects through mechanisms such as attention weighting or graph structure augmentation (Section~\ref{subsubsec:implicit-models}). 

\subsubsection{Methods that Explicitly Handle Heterophily}
\label{subsubsec:edge-type-models}

Despite the success of homophily-based solutions, \textbf{BothH}~\cite{li2023multi} challenges the homophily assumption, recognising that fake accounts engage in relation camouflage (fake accounts intentionally interact with real accounts to avoid detection), forming both homophilic and heterophilic edges. 
To address this, BothH introduces a connection classifier that differentiates between these types and adjusts its message passing strategy accordingly. By applying distinct update rules for homophilic and heterophilic edges, BothH mitigates feature contamination (unwanted feature mixing which occurs when GNNs aggregate features from heterophilic neighbours, causing real and fake node embeddings to mix, making classification harder), ensuring more accurate representations of real and fake accounts.

\textbf{CGNN}~\cite{huang2024cgnn} enhances the modelling of FAD by introducing edge-specific compatibility matrices (learnable matrices that represent the compatibility between two nodes connected by an edge, based on their types), thereby distinguishing between heterophilic and homophilic edges. Existing GNN-based approaches often fail to capture the heterogeneous nature of these interactions, treating all edges similarly despite structural differences. CGNN addresses this limitation by explicitly modelling heterogeneous compatibility through a compatibility-aware GNN encoder designed to refine message passing mechanisms based on edge types.
The CGNN encoder consists of two main components: the Edge Category Prototype Network (ECPN) and the Message Aggregation Network (MAN). ECPN is responsible for learning compatibility weight matrices for each edge by leveraging prototype embeddings and categorising interactions as fake-fake, fake-real, real-fake, or real-real. These learned compatibility matrices enable the model to capture the varying degrees of influence between different account types, ensuring that message propagation is weighted according to the network's structural properties.
Following this categorisation, MAN adapts the message aggregation process by incorporating the learned compatibility weights into the propagation mechanism. Rather than applying a uniform transformation across all edges, MAN dynamically adjusts message passing based on the identified edge categories.

\textbf{BECE}~\cite{qiao2024dispelling} enhances message passing through an Edge Confidence Evaluation (ECE) module that prunes unreliable edges, thereby improving subgraph trustworthiness (higher homophily) and refining detection accuracy.
BECE estimates edge trustworthiness by first computing the L1-difference between node embeddings to construct edge features, then uses a parameterized Gaussian network to reconstruct these features, and finally applies an MLP on the reconstructed edge embedding to output a confidence score between, enabling BECE to discard low-confidence edges before message passing. This process mitigates the adverse effects of noisy connections, ensuring that only the most trustworthy neighbour relationships contribute to node representation learning.

\subsubsection{Methods that Implicitly Handle Heterophily}
\label{subsubsec:implicit-models}

Some recent methods, such as SIRAN~\cite{zhou2023semi} and HOFA~\cite{ye2023hofa}, address the limitations of traditional GNN-based FAD by leveraging additional mechanisms, including attention or graph augmentation, to better handle heterophilic edges in graph structures.

\textbf{SIRAN}~\cite{zhou2023semi} incorporates a relation attention mechanism to reduce noisy message passing in heterophilic graphs, ensuring that features of different label nodes do not aggregate. Additionally, it utilises initial residual connections to preserve the original node features throughout training, thereby mitigating the issue of feature corruption that occurs over multiple aggregation steps.
The model uses comprehensive features, including profile, content, and graph-based features.

\textbf{HOFA}~\cite{ye2023hofa} enhances homophily by introducing a $k$-NN-based augmentation mechanism. An MLP first extracts node representations from profile, content, and graph-based features. Then, it injects additional homophilic edges via $k$-NN to reinforce local structural consistency. HOFA determines edge homophily using a threshold based on feature similarity.
In a semi-supervised scenario, where some nodes have labels and others do not, HOFA first classifies edges using node labels when available. For unlabelled nodes, it estimates edge homophily by computing the cosine similarity between node embeddings and then classifies edges based on a learnable threshold.

\textbf{HOFA}~\cite{ye2023hofa} enhances homophily through two key components: a homophily-oriented graph augmentation module that utilises an MLP to extract node representations and applies $k$-NN to add similarity-based edges, and a frequency-adaptive attention module that learns edge-level attention weights. This module is trained with a dedicated loss function to assign positive weights to homophilic edges and negative weights to heterophilic ones, thereby guiding the model to distinguish structurally consistent connections more effectively.

These methods represent a growing trend toward heterophily-aware graph learning in FAD. By modelling edge types explicitly, adjusting attention and aggregation paths, and refining the graph structure, they overcome the limitations of traditional GNNs built under homophily assumptions.

\begin{table*}[ht!]
    \centering
    \caption{Heterophily-aware methods (Section~\ref{subsec:DL:heterophilic}). Backbone models are base DL architectures. Features abbreviation: c, g, p, per (content, graph, profile, personal); nc, t (numerical/categorical, textual).} 
    % \label{tab:graph_ml_summary}
    %\renewcommand{\arraystretch}{1.2}
    \begin{tabularx}{\textwidth}{|>{\hsize=0.15\hsize}X |>{\hsize=0.12\hsize}X |>{\hsize=0.14\hsize}X |>{\hsize=0.59\hsize}X|}
        \toprule
        \textbf{Paper} & \textbf{Features} & \textbf{Backbone} & \textbf{Main Idea} \\
        \midrule
          
BothH~\cite{li2023multi} & p,c/nc,t & attention-based GNN & Learning whether each edge is homophilic or heterophilic and aggregating information accordingly in a multi-modal graph. \\
\hline
CGNN~\cite{huang2024cgnn} & p,c/nc,t & GNN & Weighting message propagation based on learned compatibility between account types. \\
\hline
BECE~\cite{qiao2024dispelling} & p,c/nc,t & GNN + Edge Confidence & Filtering unreliable edges via confidence-aware Gaussian modelling during graph learning. \\
\hline
SIRAN~\cite{zhou2023semi} & p,c/nc,t & Relational GAT & Learning heterophily-aware node representations through relation attention and residual enhancement in semi-supervised graphs. \\
\hline
HOFA~\cite{ye2023hofa} & p,c/nc,t & RGCN & Improving homophily and adaptively filtering edges through attention-based frequency control. \\
\bottomrule  
\end{tabularx}
\end{table*}

\subsection{Subgraph-Level and Community-Aware Modelling}
\label{subsec:dl:subgraph}

To move beyond simple neighbour aggregation, recent studies in FAD explore subgraph-level and community-aware modelling. These approaches extract richer structural patterns that capture the coordination, structural anomalies, and modularity of fake account behaviours. Broadly, these methods can be classified into two groups: first, methods that leverage community-aware embedding and aggregation (Sections~\ref{subsubsec:Community-AwareEmbeddingandAggregation}), and second, methods that rely on subgraphs or motifs (Section~\ref{subsubsec:Subgraph/Motif-Based Node Encoding}).

\subsubsection{Community-Aware Embedding and Aggregation}
\label{subsubsec:Community-AwareEmbeddingandAggregation}
Several methods leverage community structure to guide embedding learning, distinguishing fake accounts based on intra- and inter-community interaction patterns.

Aljohani et al.~\cite{aljohani2020bot} investigated how fake accounts integrate into the network, revealing that they are highly active within dominant communities, where they artificially inflate social engagement metrics. By applying GCNs, they demonstrated that structural patterns in OSNs played a critical role in distinguishing fake accounts from real ones. Their findings highlighted that detecting and analysing community structures can expose the influence of fake accounts. Liu et al.~\cite{liu2023Accou2Vec} introduced \textbf{Accou2Vec}, a community-aware embedding method designed to enhance FAD by improving homophily within OSNs. The method first applies Deep Autoencoder-like Non-Negative Matrix Factorisation (DANMF). This hierarchical community detection algorithm iteratively decomposes the network’s adjacency matrix to identify well-separated subgraphs with high homophily. By leveraging DANMF, Accou2Vec reduces the number of attacking heterophilic edges by segmenting the social graph into distinct communities. Once the OSN is partitioned, Accou2Vec introduces a community-aware RW strategy that distinguishes between intra-community and inter-community traces. The intra-community trace prioritises structural coherence by keeping walks within the same subgraph. In contrast, the inter-community trace incorporates a penalty term discouraging traversal across communities. 

Following a similar approach, \textbf{Bot2Vec}~\cite{pham2022bot2vec} focuses on learning embeddings that preserve intra-community structures. Instead of aggregating features from neighbours, it constructs node representations using RWs biased to stay within the same community, identified by the Louvain algorithm. This idea helps identify fake account clusters that exhibit dense internal connectivity but have limited interactions with real accounts.
Bot2Vec modifies the standard Node2Vec~\cite{grover2016node2vec} transition probability by incorporating an intra-community bias, prioritising transitions within the same community while penalising those outside it.

\subsubsection{Subgraph/Motif-Based Node Encoding}
\label{subsubsec:Subgraph/Motif-Based Node Encoding}

Another line of work aims to improve node representations by capturing structural motifs and encoding information from local subgraphs. \textbf{SEGCN}~\cite{liu2024segcn} is a subgraph encoding-based GCN that enhances the expressiveness of GCNs by encoding each node based on its induced subgraphs rather than relying solely on its immediate neighbours, which enables the model to recognise critical network motifs, such as cycles and triangles, often indicative of coordinated fake account behaviours. SEGCN constructs induced subgraphs for each node using an RW-based sampling technique, ensuring that subgraphs remain informative and computationally efficient. A GCN-based encoder processes these subgraphs, enabling the model to aggregate node information from both immediate and distant neighbours, thereby enhancing expressiveness.

\textbf{BSG4Bot}~\cite{miao2024bsg4bot} selectively constructs subgraphs biased toward homophilic connections to reduce noise and improve scalability in large graphs. BSG4Bot trains an MLP classifier on profile and content features to estimate node similarities and integrates these with Personalised PageRank (PPR) scores to construct subgraphs favouring homophilic neighbours. These biased subgraphs are then processed using a heterogeneous GNN with attention. Training on these subgraphs instead of the entire network, BSG4Bot significantly reduces memory consumption and training time while preserving or even surpassing the accuracy of state-of-the-art methods.

\begin{table*}[ht!]
    \centering
    \caption{Subgraph-level and community-aware modelling based methods (Section~\ref{subsec:dl:subgraph}). Backbone models are base DL architectures. Features abbreviation: c, g, p, per (content, graph, profile, personal); nc, t (numerical/categorical, textual).} 
    % \label{tab:graph_ml_summary}
    %\renewcommand{\arraystretch}{1.2}
    \begin{tabularx}{\textwidth}{|>{\hsize=0.15\hsize}X |>{\hsize=0.12\hsize}X |>{\hsize=0.14\hsize}X |>{\hsize=0.59\hsize}X|}
        \toprule
        \textbf{Paper} & \textbf{Features} & \textbf{Backbone} & \textbf{Main Idea} \\
        \midrule
          
Aljohani et al.~\cite{aljohani2020bot} & g/nc & GCN & Deploying a featureless GCN using only graph structure. \\
\hline
Accou2Vec~\cite{liu2023Accou2Vec} & g/n & Deep Autoencoder-like & Learning structure-aware embeddings via community-constrained walks.\\
\hline
Bot2vec~\cite{pham2022bot2vec}  & g/n & DeepWalk  & Learning account representations by combining intra-community and neighbourhood-aware RWs. \\
\hline
Bot2Vec~\cite{liu2024segcn} & p,c/nc,t & GCN & Enhancing graph convolution by encoding subgraph structures. \\
\hline
BSG4Bot~\cite{miao2024bsg4bot} & p,c/nc,t & GCN & Training on biased subgraphs with enhanced homophily using node similarity and relational structure. \\
\bottomrule 
\end{tabularx}
\end{table*}

\subsection{Multi-Relational Graphs}
\label{subsec:DL-Het-MR}
OSNs can be modelled as graphs with multiple relations (multi-relational graph) and node types, encompassing diverse entities and interaction types. In multi-relational graphs, edges can represent various interactions such as friendships, follows, retweets, and likes.
To effectively capture this complexity, learning methods designed for multi-relational graphs and multiple node types have been developed, providing enhanced representation for FAD.

% \ali{subsection  and subsubsection titles changed} \anz{I commented out the last part of the paragraph above. Since it's a short section, that's not necessary.}\ali{thank you.}
% \anz{I also updated the titles for subsection and subsubsections. Please check of they make sense and if yes make changes wherever necessary.}\ali{done.}

\subsubsection{Single Node Type}
% \ali{Dear @Ahad, I suggest \textbf{Single} node type instead of \textbf{One}}
\label{subsubsec:Multi-Relational Graph Approaches for Homogeneous Networks}

Schlichtkrull et al.\cite{schlichtkrull2018modeling} introduced Relational Graph Convolutional Networks (RGCNs) to enable graph convolution on multi-relational graphs by incorporating relation-specific transformations. This framework has served as a foundation for modelling multi-relational OSNs.
\textbf{BotRGCN}~\cite{feng2021botrgcn} was one of the first models to apply RGCNs by constructing a multi-relational graph. It integrates numerical, categorical, and textual profile features, along with textual content features from account-generated text, into a unified relational embedding. This embedding is then processed through RGCNs to identify fake accounts that either mimic real accounts or fake ones. Recently, using BotRGCN, Reiche et al.~\cite{reiche2024integrating} incorporated co-retweet and co-hashtag relations, demonstrating that these coordinated activities can serve as strong indicators of fake accounts.

Building upon this line of work on RGCN, Tzoumanekas et al.~\cite{tzoumanekas2024graph} proposed \textbf{DFG-NAS}, a Neural Architecture Search (NAS) method for FAD. DFG-NAS adapts NAS to RGCNs by automatically searching for optimal configurations of propagation and transformation functions using an evolutionary algorithm (an optimisation algorithm inspired by the process of natural selection in biological evolution). This approach enhances detection performance while avoiding the limitations of fixed GNN architectures.

As the next step, in \textbf{BotRGA}~\cite{wang2024botrga}, Relational Graph Aggregation (RGA) was introduced to capture direct and indirect neighbour influences.
Using a relation-type-based attention mechanism referred to in this research as a semantic fusion network, BotRGA integrates information from different relations.
This design mitigates the limitations of transductive learning and enhances generalisation to previously unseen nodes.

\textbf{TCAE-DL-RGCN}~\cite{du2024tcae} is built on the foundation of RGCN. It aims to enhance the expressiveness of node representations, particularly addressing the over-smoothing problem that arises in deep GNNs. To tackle this, the method employs dual RGCN branches: one is a standard RGCN, and the other is augmented with a self-attention mechanism modulated by a controllable temperature parameter (a scalar in the softmax function that adjusts the sharpness of the attention distribution). This temperature-controlled attention prevents dominance by a few nodes and helps maintain diversity in node representations. The outputs of both branches are combined to obtain richer embeddings.

\textbf{Bot-MGAT}~\cite{alothali2022bot} introduces a multi-view attention-based architecture designed specifically for transfer learning, addressing the scarcity of labelled data and domain shifts common in FAD. It constructs multiple graph views, each representing a different relationship type (e.g., follower/following, interactions such as replies or mentions), and applies a separate GAT module to each view. These view-specific embeddings are aggregated via a weighted sum (using a fixed parameter) to produce the final node representation. Unlike TCAE-DL-RGCN, Bot-MGAT only utilises profile features (both numerical and categorical), deliberately avoiding textual features such as tweets or descriptions to reduce complexity and improve scalability.

Peng et al.~\cite{peng2024coarse} proposed a semi-supervised method called Coarse-to-Fine Label Propagation (\textbf{LP-CF}) to address the challenge of limited labelled data. This approach is based on two core components: Hybridised Representation Models over Multi-Relational Graphs (HR-MRG) and a tailored label propagation strategy that adapts to the difficulty level of each sample.
The HR-MRG framework employs two models. The first, referred to as the basic model, is trained on all labelled accounts and captures general patterns. It is then used to generate confidence scores for unlabelled accounts. High-confidence predictions are categorised as easy to classify (referred to as the coarse group). In contrast, low-confidence predictions are marked as hard to classify or confusing (referred to as the fine group).
To better represent these challenging cases, a second model, known as the rectified model, is trained solely on the subset of labelled accounts that the basic model struggles to classify with high confidence. This specialisation enables the rectified model to capture subtle patterns missed by the basic model. Notably, the rectified model is not trained directly on unlabelled data but is used to generate high-quality representations for the hard-to-classify samples.

\subsubsection{Multiple Node Types}
\label{subsubsec:Heterogeneous Graph Approaches with Node/Edge Diversity}

While most works primarily focus on multi-relational graphs, another line of research explores multiple node types in which different nodes, such as accounts, posts, and communities, possess distinct attributes.
\textbf{SybilFlyover}\cite{li2022sybilflyover} utilises graphs with multiple node and edge types for FAD. It models OSNs as directed graphs consisting of different node types, including account and tweet nodes, and incorporates various edge types such as follow, retweet, and like relations. This structure enables the model to capture account–account and account–tweet interactions, addressing the limitations of previous structure-based approaches that relied solely on single-type node representations.
SybilFlyover integrates textual content features into the detection pipeline, combining content-aware embeddings with structural information to create more informative representations of account behaviour. To process the resulting multi-type graph, SybilFlyover utilises the Heterogeneous Graph Transformer (HGT)\cite{hu2020heterogeneous}, a neural architecture designed explicitly for graphs with multiple node and edge types. HGT extends the transformer mechanism by using type-specific attention and message passing functions, allowing the model to weigh the importance of different node and relation types during aggregation.

Building on the idea of using different node types, \textbf{BGSRD}~\cite{guo2021social} constructs a multi-node-type graph by adding tweets (referred to as documents) and words as nodes with new types. 
The model assigns weights to document–word edges using TF-IDF scores, indicating the importance of each word within a tweet. Word–word edges are weighted using Positive Pointwise Mutual Information (PPMI), which captures the strength of word co-occurrence across the dataset. It combines embeddings generated by BERT (a pre-trained language model) with a GCN to capture both semantic and graph-based features. Tweets are labelled based on whether they originate from fake or real accounts, enabling content-based FAD.

\textbf{GNNRI}~\cite{li2024gnnri} is a framework that leverages graphs with multiple node and relation types to incorporate explicit and implicit relationships among accounts, tweets, comments, and hashtags. It models explicit relations, such as retweeting and commenting, alongside implicit ones, such as accounts posting tweets with the same hashtags or exhibiting similar behavioural patterns even without direct interactions. These relationships are represented through meta-structures, specifically meta-paths and meta-graphs.
GNNRI introduces two core components to exploit these structures: a Relation-based Self-Attention Layer (RSL) and an Implicit-Connection Convolutional Layer (ICL). The RSL computes attention over different meta-structures, allowing the model to aggregate neighbour information under specific relational contexts. The ICL constructs a similarity matrix in which each entry reflects the semantic closeness between two accounts, presented as a function of the number of meta-path or meta-graph instances connecting them. This matrix is then fed into a GCN to propagate information across accounts, including those without direct links.

\begin{table*}[ht!]
    \centering
    \caption{Multi-Relational Graphs (Section~\ref{subsec:DL-Het-MR}). Backbone models refer to the base DL architectures used. Features are abbreviated as follows: c, g, p, per (content, graph-based, profile, personal)/ nc, t (numerical or categorical, textual).} 
    % \label{tab:graph_ml_summary}
    %\renewcommand{\arraystretch}{1.2}
    \begin{tabularx}{\textwidth}{|>{\hsize=0.15\hsize}X |>{\hsize=0.12\hsize}X |>{\hsize=0.14\hsize}X |>{\hsize=0.59\hsize}X|}
        \toprule
        \textbf{Paper} & \textbf{Features} & \textbf{Backbone} & \textbf{Main Idea} \\
        \midrule

BotRGCN~\cite{feng2021botrgcn} & p,c/nc,t & RGCN & Combining multi-modal features with relational graph convolution. \\
\hline
Reiche et al.~\cite{reiche2024integrating} & p,c/nc,t & RGCN & Integrating higher-order behavioural relations like co-retweet and co-hashtag into graph learning. \\
\hline
DFG-NAS~\cite{tzoumanekas2024graph} & p,c/nc,t & RGCN & Optimising RGCN message passing architecture using neural architecture search. \\
\hline
BotRGA~\cite{wang2024botrga} & p,c/nc,t & Relational Graph Aggregation & Aggregating neighbourhood information through relation-aware and semantically weighted graph aggregation. \\
\hline
TCAE-DL-RGCN~\cite{du2024tcae} & p,c/nc,t & RGCN & Fusing features with relational attention and temperature control. \\
\hline
Bot-MGAT~\cite{alothali2022bot} & p/nc & GAT + RGCN & Combining multi-view graph attention and transfer learning across domains (accounts' interests). \\
\hline
LP-CF~\cite{peng2024coarse} & p,c/nc,t & Multi-Relational GNN & Refining label propagation over hybrid GNN representations from multi-relational graphs. \\
\hline
SybilFlyover~\cite{li2022sybilflyover} & p,c/nc,t & HGT & Modelling accounts and textual contents in a directed graph and learning semantic and structural relations using prompt-enhanced representations. \\
\hline
BGSRD~\cite{guo2021social} & c/t & GCN & Combining BERT embeddings with graph-based co-occurrence learning. \\
\hline
GNNRI~\cite{li2024gnnri} & p,c/nc,t & GAT+GCN & Modelling explicit and implicit connections between accounts in a graph with relation-aware attention and convolution. \\
\bottomrule  
\end{tabularx}
\end{table*}

\subsection{Incorporating Content Features}
\label{subsec:Incorporating Content Features}

The growing sophistication of fake accounts has driven the adoption of textual and visual features analysis from profile and content data.  

A subset of methods focuses on understanding the deeper semantics of textual content, such as topics, emotions, and behavioural consistency over time.
For example, \textbf{BIC}~\cite{lei2022bic} integrates textual features and graph structure through interactive learning, allowing the two to update each other during training.
BIC introduces two main innovations: a text–graph interaction module that enables deep, similarity-based mutual updates between text and graph representations and a semantic consistency module that captures unusual changes in an account's textual content over time. The model first encodes textual data using RoBERTa (a pre-trained language model) and graph data using RGCN. It then performs multiple rounds of cross-modal interaction to exchange information between modalities. Attention weights from the text module are used to derive semantic consistency between features, highlighting temporal inconsistencies in account behaviour. Finally, BIC fuses the text, graph, and consistency representations to perform FAD.

\textbf{ETS-MM}\cite{li2025ets} improves FAD by leveraging large language models. It focuses on two main aspects of account behaviour: the topics they talk about most often and the emotions they tend to express. The main idea is to go beyond just reading the text and capture patterns in accounts’ interests and emotional expressions. ETS-MM combines this deep understanding of an account's textual information with additional information, such as follower counts or verification status, and utilises a graph structure using a GNN.

Another set of methods develops robust fusion strategies to integrate textual, structural, and profile modalities into unified account representations.
Goyal et al.~\cite{goyal2023detection} proposed a multi-modal DL framework that integrates CNNs for visual content (e.g., profile and banner images), LSTM networks for textual data (e.g., tweets), and a GCN for modelling OSN structure. Their approach explicitly combines content and graph-based features to enhance FAD.

\textbf{BotSAI}~\cite{gong2024enhancing} is a framework designed to integrate and align diverse account information, including profile, textual content, and graph-based features within a multi-relational network. Each of these information types is treated as a separate modality. The core idea is to project each modality into two subspaces: an invariant subspace that captures common information across modalities, and a specific subspace that preserves modality-unique characteristics. This dual-space design enables the model to jointly exploit shared behavioural patterns and modality-specific signals for robust representation learning.
Each modality is encoded using a modality-specific encoder: an MLP for metadata features, a pre-trained RoBERTa model for textual content, and a custom graph encoder based on a local relational graph transformer (an extended version of Relational Graph Transformer (RGT)) for capturing structural patterns in the multi-relational social graph. The resulting feature vectors are projected into both invariant and specific subspaces. A multi-head self-attention mechanism is then employed to fuse these representations, producing a unified embedding for each account, which is subsequently used for FAD.

A complementary approach explores self-supervised learning to generate generalisable multi-modal embeddings for FAD without relying on labelled data. For example, \textbf{SATAR}~\cite{feng2021satar} addresses this by pre-training on a follower count prediction task. The model computes the embedding of textual content and profile features. For textual features, a hierarchical bidirectional RNN with an attention module captures both tweet-level and word-level semantics. Neighbourhood information is aggregated without applying GNN by combining followers’ semantic embeddings through aggregation and averaging followers' profile embeddings. These modalities are fused via an aggregator function, which learns pairwise correlations (e.g., how profile features interact with linguistic patterns in textual features) to produce a unified representation. After pre-training on unlabelled data, the fine-tuning process is on FAD with limited labels.

\begin{table*}[ht!]
    \centering
    \caption{Incorporating Content Features (Section~\ref{subsec:Incorporating Content Features}). Backbone: base DL models. Features: c, g, p, per (content, graph, profile, personal); nc, t (numerical/categorical, textual).} 
    % \label{tab:graph_ml_summary}
    %\renewcommand{\arraystretch}{1.2}
    \begin{tabularx}{\textwidth}{|>{\hsize=0.15\hsize}X |>{\hsize=0.12\hsize}X |>{\hsize=0.14\hsize}X |>{\hsize=0.59\hsize}X|}
        \toprule
        \textbf{Paper} & \textbf{Features} & \textbf{Backbone} & \textbf{Main Idea} \\
        \midrule
          
BIC~\cite{lei2022bic} & p,c/t & RGCN & Exchanging information between text and graph modalities and modelling semantic consistency across tweets. \\
\hline
ETS-MM~\cite{li2025ets} & p,c/nc,t & GNN & Combining enhanced textual features with metadata and social relations using LLMs and GNNs. \\
\hline
Goyal et al.~\cite{goyal2023detection} & p,c/nc,t & GCN & Combining multi-modal features. \\
\hline
BotSAI~\cite{gong2024enhancing} & p,c/nc,t & RGT & Fusing multi-modal invariant and specific account representations. \\
\hline
SATAR~\cite{feng2021satar} & p,c,g/nc,t & Encoder + GNN & Predicting follower count to learn self-supervised multi-modal embeddings with feature fusion. \\
\bottomrule 
\end{tabularx}
\end{table*}

\subsection{Contrastive Learning Based Methods}
\label{subsec:DL-CL}

Contrastive Learning (CL) has emerged as a powerful paradigm for FAD, offering robust and label-efficient representation learning in noisy, heterophilic OSNs. These CL-based models can be categorised into three key classes: supervised CLs (Section~\ref{subsubsec:DL-CL-supervised}); unsupervised and self-supervised CLs (Section~\ref{subsubsec:DL-CL-unsupervised}); and task-customised and advanced CLs (Section~\ref{subsubsec:dl-cl-third}) include
CL methods that are specifically adapted to the unique challenges of FAD, such as class imbalance, behavioural diversity, and structural complexity in OSNs. 

\subsubsection{Supervised Contrastive Learning}
\label{subsubsec:DL-CL-supervised}

As discussed earlier, heterophilic edges degrade GNN performance by causing undesired feature mixing between different classes.
To counter this, \textbf{BotSCL}~\cite{wu2023botscl} employs supervised CL, which explicitly pulls together representations of nodes from the same class while pushing apart those from different classes. At the core of BotSCL is a channel-wise encoder that addresses the heterophily issue in message passing. Unlike conventional GNNs that uniformly aggregate all neighbour features, this encoder processes each feature channel independently. It computes per-feature attention coefficients for each neighbour, allowing the model to extract useful information from homophilic neighbours while discriminating against misleading information from heterophilic ones.
BotSCL applies its contrastive objective across two augmented graph views to enhance the robustness of representations. The first augmentation, class-aware node shuffling, randomly swaps feature vectors between nodes of the same class, enforcing invariance to neighbourhood structure while preserving label semantics. The second, edge removal, perturbs the graph topology to encourage generalisation beyond the original connections. These views are passed through the encoder, and the resulting node embeddings are optimised using a cross-view supervised contrastive loss.

Building on this heterophily-aware design, CACL~\cite{chen2024cacl} shifts the focus to the community-level structure of OSNs. Rather than treating all node pairs equally, CACL identifies same-label nodes from different communities as hard positives, since they are semantically similar despite being distant in the graph. Conversely, different-class nodes within the same community are treated as hard negatives, as their closeness can be misleading.
To identify these hard samples, CACL employs a GNN to compute node representations and applies a clustering model to detect communities. The learned node similarities are used to iteratively group accounts, forming subgraphs where nodes exhibit high representation similarity. This process promotes the formation of tighter representation clusters within communities and greater separation across them, reflecting the natural modularity of OSNs. Based on the detected communities, hard positive and negative samples are dynamically mined during training.
These samples are then used in a supervised CL that pulls same-class nodes from different communities closer and pushes different-class nodes within the same community apart. CACL enhances the model's ability to learn node embeddings that serve as inputs for FAD.

\subsubsection{Unsupervised and Self-Supervised Contrastive Learning}
\label{subsubsec:DL-CL-unsupervised}

\textbf{BotDCGC}~\cite{wang2024unsupervised} proposes a fully unsupervised framework that combines graph autoencoding, CL, and deep clustering in a unified pipeline. Its key novelty is jointly learning node embeddings and the assignment of nodes to clusters without label supervision.
BotDCGC begins by encoding diverse account attributes, including numerical, categorical, and textual profile features, as well as textual content features, using Bi-LSTM and pre-trained models. These representations are then refined by a graph attention encoder that integrates structural information from multi-hop neighbourhoods, extending beyond standard GAT by considering broader graph connectivity. An inner product decoder reconstructs account relationships to preserve graph structure.
To make the embeddings more discriminative, BotDCGC applies CL, encouraging neighbouring nodes to be closer in the embedding space while pushing distant ones apart.

\textbf{BotCL}~\cite{li2024botcl} is a graph CL framework designed to enhance robustness against structural manipulation in OSNs. In comparison to previous methods, BotCL distinguishes itself by directly generating graph views based on the graph structure through two operations: node dropping and edge perturbation. These augmentations simulate realistic OSN dynamics, such as account suspension and changes in follow relationships, generating multiple graph views that reflect possible manipulation states.
Each view is encoded using an RGCN, which effectively captures directionality and relation types in the graph, including following and follower relation types. BotCL then applies a contrastive loss to maximise the consistency between representations of the same node across different augmented views while distinguishing them from other nodes. By combining profile, content, and graph-based features in the input, BotCL learns node representations through structure-aware.

\textbf{SeBot}~\cite{yang2024sebot} uses structural entropy minimisation to uncover the hidden hierarchical organisation of a network by constructing encoding trees that reflect both local and global structural patterns. By aligning multiple structural views through CL, SeBot aims to mitigate feature contamination caused by fake accounts attaching themselves to real account subgraphs. Its novelty lies in combining structural entropy with CL to produce node representations at multiple levels of granularity, capturing both the fine structure of local neighbourhoods and the broader context of global communities.
SeBot compresses the graph by merging nodes into communities that minimise structural uncertainty and then reconstructs node-level embeddings enriched with high-order structural information.

\subsubsection{Task-Customised and Advanced Contrastive Learning}
\label{subsubsec:dl-cl-third}

\textbf{CBD}~\cite{zhou2023detecting} proposes a two-stage CL framework: an unsupervised pre-training phase to capture general structural patterns from large-scale unlabelled data, followed by a semi-supervised fine-tuning phase that adapts to emerging fake behaviours using only a few labelled examples. The pre-training stage employs CL to derive generalisable node representations, while the fine-tuning phase enhances prediction reliability with limited annotations. This design enables CBD to achieve effective real-time FAD even in few-shot settings.

A more sophisticated approach, \textbf{SeGA}~\cite{chang2024sega}, introduces a preference-aware CL framework by leveraging account-specific posting behaviours rather than relying solely on structural or profile features. The core idea is to use large language models to extract each account’s preferred topics and emotions from their recent textual content, summarising them as topic-emotion pairs. These preferences are embedded into natural language prompts that serve as pseudo-labels in a CL objective, aligning each account’s embedding with their behavioural signature while distinguishing them from others.
SeGA encodes accounts and lists as nodes within a network and employs an RGT to model the diverse interactions between them. The framework consists of three stages: feature encoding, preference-aware contrastive pre-training, and fine-tuning for classification.

\textbf{DGBot}~\cite{xu2024dgbot} addresses the over-globalisation problem in graph transformers, where attention is overly distributed across distant nodes, causing the model to overlook important local structures. DGBot first partitions the graph into clusters and employs two Transformer modules to extract intra-cluster (local) and inter-cluster (global) features to mitigate this. It also applies an RGCN to capture fine-grained, relation-specific neighbour information. Finally, DGBot uses CL to align and balance the representations from CGA and RGCN modules, enhancing its ability to detect fake accounts that blend into real account communities by preserving local and global cues.

\begin{table*}[ht!]
    \centering
    \caption{Contrastive learning methods (Section~\ref{subsec:DL-CL}). Backbone: base DL models. Features: c, g, p, per (content, graph, profile, personal); nc, t (numerical/categorical, textual).} 
    % \label{tab:graph_ml_summary}
    %\renewcommand{\arraystretch}{1.2}
    \begin{tabularx}{\textwidth}{|>{\hsize=0.15\hsize}X |>{\hsize=0.12\hsize}X |>{\hsize=0.14\hsize}X |>{\hsize=0.59\hsize}X|}
        \toprule
        \textbf{Paper} & \textbf{Features} & \textbf{Backbone} & \textbf{Main Idea} \\
        \midrule
          
BotSCL~\cite{wu2023botscl} & p,c/nc,t & CL+ Transformer & Combining homophilic and heterophilic information through supervised and channel-aware graph encoding. \\
\hline
CACL~\cite{chen2024cacl} & p,c/nc,t & CL+(GAT, SAGE, HGT) & Training on hard positive and hard negative samples mined from community structure. \\
\hline
BotDCGC~\cite{wang2024unsupervised} & p,c/nc,t & GAT+CL & Clustering by jointly optimising graph-based embeddings with an unsupervised deep clustering. \\
\hline
BotCL~\cite{li2024botcl} & p,c/nc,t & RGCN & Learning robust node embeddings through on augmented directed graphs with semantic, attribute, and graph-based features. \\
\hline
SeBot~\cite{yang2024sebot} & p,c/nc,t & CL & Aligning hierarchical and relational representations through entropy-guided multi-view. \\
\hline
CBD~\cite{zhou2023detecting} & p,c/nc,t & GCN+CL & Combining contrastive pre-training and consistency-aware fine-tuning for few-shot FAD. \\
\hline
SeGA~\cite{chang2024sega} & p,c,g/nc,t & RGT + CL & Leveraging topic-emotion preferences from posts as pseudo-labels. \\
\hline
DGBot~\cite{xu2024dgbot} & p,c/nc,t & RGCN + Transformer & Combining global cluster-level attention and local relation-aware features. \\
\bottomrule 
\end{tabularx}
\end{table*}

\subsection{Reinforcement Learning Based Methods}
\label{subsec:DL-RL}

RL has emerged as a powerful tool for enhancing FAD by introducing dynamic adaptability into graph-based methods.
RL allows models to adjust graph sampling strategies, optimise network depth, and tailor the detection process in response to the evolving tactics of fake accounts, thereby improving robustness and generalisation.

\textbf{CARE-GNN}~\cite{dou2020enhancing} combines a supervised similarity measure, an adaptive neighbour selector, and a relation-aware aggregator. The method first uses a label-aware MLP to compute similarity scores between nodes. Based on these scores, it selects the most informative neighbours using a dynamically adjusted threshold during training. This threshold is tuned using a simple greedy strategy inspired by RL, where the model increases or decreases the threshold depending on whether the selected neighbours become more similar across epochs. Finally, the chosen neighbours are aggregated across multiple relations using the learned thresholds as weights. 

\textbf{RoSGAS}~\cite{yang2023rosgas} combines self-supervised learning with an RL-based GNN architecture search. The key idea is to treat FAD as a personalised subgraph classification task, where the model adaptively selects both the subgraph size and the number of GNN layers for each individual account. This approach addresses the challenges posed by evolving camouflage strategies and the scarcity of labelled data.
The process begins by constructing a network from social data, incorporating accounts, shared textual content, hashtags, and their interactions. For each target account, RoSGAS extracts a local subgraph for label prediction. It utilises two RL agents to optimise the GNN design: one selects the neighbourhood depth (i.e., the number of hops to include), and the other determines the number of GNN layers. These decisions are made on a per-account basis, allowing the model to tailor its architecture to the specific local structure of each subgraph.

\begin{table*}[ht!]
    \centering
    \caption{Reinforcement learning (Section~\ref{subsec:DL-RL}). Backbone: base DL models. Features: c, g, p, per (content, graph, profile, personal); nc, t (numerical/categorical, textual).} 
    % \label{tab:graph_ml_summary}
    %\renewcommand{\arraystretch}{1.2}
    \begin{tabularx}{\textwidth}{|>{\hsize=0.15\hsize}X |>{\hsize=0.12\hsize}X |>{\hsize=0.14\hsize}X |>{\hsize=0.59\hsize}X|}
        \toprule
        \textbf{Paper} & \textbf{Features} & \textbf{Backbone} & \textbf{Main Idea} \\
        \midrule
          
CARE-GNN~\cite{dou2020enhancing} & p,c,g/nc,t & GCN + CNN + RL & Detecting camouflaged fraudsters via multi-view feature fusion and RL-based threshold tuning. \\
\hline
RoSGAS~\cite{yang2023rosgas} & p,c/nc,t & RL-based GNN & learning subgraph size and GNN architecture adaptively using RL and self-supervised learning. \\
\bottomrule  
\end{tabularx}
\end{table*}

\subsection{Temporal Methods}   
\label{subsec:DL-temporal}

Many FAD methods assume static account behaviour, which limits their effectiveness in capturing the adaptive nature of fake accounts. To overcome this challenge, some methods incorporate temporal dynamics into the detection process, enabling models to track behavioural changes over time and uncover patterns indicative of manipulation or automation.
These approaches combine temporal activity information with relational modelling, capturing both behaviour and accounts structure role evolution. 
By representing the OSN as a sequence of graph snapshots, they detect fake accounts based on their changing patterns of connections and content.

\textbf{BotWS}\cite{qiao2023social} models how an account's interests change over time based on its posting history. The key idea is that, although modern fake accounts can generate content that looks similar to real account posts, they often fail to show the stable and personal posting patterns seen in real accounts. 
To capture this difference, BotWS splits each account's posts into time windows and uses multi-head attention to track how interests change across them over time. These patterns are then encoded into the account representation. 
This design enables the model to identify unstable or inconsistent interest changes, which, as claimed in the paper, are more commonly observed in fake accounts.
To include relational information, BotWS builds a multi-relational graph where nodes represent accounts and edges capture social connections such as follower and following links. Finally, a simple heterogeneous graph neural network (Simple-HGN)\cite{lv2021we} is applied to model both the types of edges and the influence between accounts, combining behavioural and structural information for FAD.

\textbf{BotDGT}~\cite{he2024dynamicity} presents a dynamic graph-based FAD framework that captures both structural and temporal patterns in OSNs. The key idea is to model OSNs as dynamic graphs composed of sequences of snapshots over time rather than relying on a single, static view. This dynamic representation enables the model to track how account behaviour evolves and detect fake accounts that adapt over time to evade detection.
BotDGT's core novelty lies in its dual-module architecture, which captures both the structural and temporal characteristics of OSNs. The structural module employs a message passing mechanism based on multi-head dot-product attention. Inspired by transformer architectures, this mechanism is adapted to operate on graph snapshots, where it learns node representations by aggregating information from neighbours, assigning different importance to each based on the graph structure at each time step.
The temporal module also adopts a transformer-style self-attention mechanism but focuses on modelling the evolution of individual nodes across time.

These methods reflect the field’s shift from detecting simple temporal patterns to capturing complex behavioural changes over time. As fake accounts become increasingly adaptive, temporal modelling is essential for effective detection.

\begin{table*}[ht!]
    \centering
    \caption{Temporal methods (Section~\ref{subsec:DL-temporal}). Backbone: base DL models. Features: c, g, p, per (content, graph, profile, personal); nc, t (numerical/categorical, textual).} 
    % \label{tab:graph_ml_summary}
    %\renewcommand{\arraystretch}{1.2}
    \begin{tabularx}{\textwidth}{|>{\hsize=0.15\hsize}X |>{\hsize=0.12\hsize}X |>{\hsize=0.14\hsize}X |>{\hsize=0.59\hsize}X|}
        \toprule
        \textbf{Paper} & \textbf{Features} & \textbf{Backbone} & \textbf{Main Idea} \\
        \midrule
          
BotWS~\cite{qiao2023social} & p,c/nc,t & HGN & Modelling attention-based interest changes across posting windows and account relationships. \\
\hline
BotDGT~\cite{he2024dynamicity} & g/nc & Graph Transformer & Capturing evolving structural patterns over time with dynamic graph transformers. \\
\bottomrule 
\end{tabularx}
\end{table*}

\subsection{Mixture of Experts}
\label{subsec:DL-MOE}

Mixture of Experts (MoE) is an ML technique where multiple specialised neural networks, called experts, collectively handle a task. Each expert processes inputs relevant to its specialisation, and the gating mechanism dynamically weights their contributions to produce the final output.

Extending the MoE concept, Lu et al.~\cite{lu2024adaptive} proposed \textbf{DSBD}, a domain-aware method for FAD across multiple interest categories (domains), such as politics or sports. Newer fake accounts often mimic real accounts by interacting with content from various domains, making them more challenging to detect.
To model this behaviour, DSBD estimates an account’s domain interests without relying on labelled training data for each domain. Instead, it uses a large pre-trained language model to analyse the account’s posts and estimate the likelihood of interest in each domain. These probability scores, known as soft domain labels, reflect mixed interests. DSBD then uses an MoE structure, where each expert specialises in one domain and is implemented using an RGT. To combine the expert outputs, DSBD introduces a domain gate as a small neural network that computes the importance of each domain expert by considering both the soft domain labels and the account's overall text content. The final account embedding is then passed to an MLP for classification.

\textbf{BotMoE}~\cite{liu2023botmoe} introduces a community-aware MoE framework for FAD. The central novelty of BotMoE lies in its adaptive expert selection mechanism, which enables the model to adjust dynamically to the diversity of account behaviours across communities. BotMoE processes each account through separate expert modules for each modality. These modalities include numerical and categorical profile features, textual profile and content features, and graph structure. For each modality, a gating network analyses the input and activates a small set of experts using top-$k$ selection. These experts are shared across all accounts and are not tied to specific predefined communities. Instead, they dynamically learn to specialise in modelling account behaviours from different communities during training. This adaptive expert selection enables BotMoE to generalise effectively across diverse and unseen account communities.
The method operates in multiple stages: raw features are encoded; gating networks assign inputs to experts; expert outputs are fused via a transformer-based module, which also computes a consistency matrix over the attention map to detect feature manipulation. The final fused representation is passed to a classifier for FAD.

\begin{table*}[ht!]
    \centering
    \caption{Mixture of experts methods (Section~\ref{subsec:DL-MOE}). Backbone: base DL models. Features: c, g, p, per (content, graph, profile, personal); nc, t (numerical/categorical, textual).} 
    % \label{tab:graph_ml_summary}
    %\renewcommand{\arraystretch}{1.2}
    \begin{tabularx}{\textwidth}{|>{\hsize=0.15\hsize}X |>{\hsize=0.12\hsize}X |>{\hsize=0.14\hsize}X |>{\hsize=0.59\hsize}X|}
        \toprule
        \textbf{Paper} & \textbf{Features} & \textbf{Backbone} & \textbf{Main Idea} \\
        \midrule
         
DSBD~\cite{lu2024adaptive} & p,c/nc,t & RGT & Leveraging domain-aware learning through MoE. \\
\hline
BotMoE~\cite{liu2023botmoe} & p,c/nc,t & RGCN & Leveraging modality-specific expert networks and community-aware fusion. \\
\bottomrule 
\end{tabularx}
\end{table*}

\subsection{Federated Learning}
\label{subsec:federated_learning}
Federated learning is known for offering data privacy by training models across multiple participants (called clients) who each hold their own data and perform local training, without sharing raw data. This advantage is important in FAD tasks where account privacy must be protected.

Peng et al.~\cite{peng2022domain} proposed \textbf{DA-MRG}, a Domain-Aware Multi-Relational Graph model with federated learning for FAD. The model constructs multi-relational graphs using profile and content features and relationships and incorporates domain-aware classifiers that recognise behavioural differences across accounts belonging to different domains. Here, a domain refers to an account-level attribute that indicates the context or subject area an account primarily engages with  (e.g., politics or business). By leveraging this domain-specific information, the model adapts its classification strategy to better distinguish between fake and real accounts within each domain. To address privacy concerns, DA-MRG is extended with a federated learning framework, enabling collaborative training without the need to share sensitive raw data directly. 

Wang et al.~\cite{wang2024fedkg} proposed \textbf{FedKG}, which improves FAD by combining federated learning with a technique called knowledge distillation. In this method, each federated learning client trains a model using its own data, while a central server learns from all clients to generate synthetic examples that capture shared patterns. These examples are then sent back to help improve each client's model. This setup enables collaborative learning without sharing raw data, resulting in higher accuracy and fewer communication rounds, even when the data across clients is significantly different.

Yang et al.~\cite{yang2023fedack} proposed \textbf{FedACK}, a non-graph-based framework that combines federated learning, adversarial learning, and knowledge distillation to address FAD across different languages and model architectures. Unlike graph-based methods, FedACK operates on account metadata and textual content. Each federated learning client, representing a platform or device with private data, trains a personalised model while contributing to a globally consistent feature space. 
To handle multilingual data, FedACK integrates a cross-lingual mapping module that projects texts from different languages into a shared context space using a Transformer-based encoder-decoder and adversarial training. Additionally, it employs a GAN-based global generator to capture the overall data distribution, a multi-stage adversarial and contrastive training strategy to align local and global feature spaces, and two discriminators (global and local) to support model customisation and reduce divergence across clients.

\begin{table*}[ht!]
    \centering
    \caption{Federated learning methods (Section~\ref{subsec:federated_learning}). Backbone: base DL models. Features: c, g, p, per (content, graph, profile, personal); nc, t (numerical/categorical, textual).} 
    % \label{tab:graph_ml_summary}
    %\renewcommand{\arraystretch}{1.2}
    \begin{tabularx}{\textwidth}{|>{\hsize=0.15\hsize}X |>{\hsize=0.12\hsize}X |>{\hsize=0.14\hsize}X |>{\hsize=0.59\hsize}X|}
        \toprule
        \textbf{Paper} & \textbf{Features} & \textbf{Backbone} & \textbf{Main Idea} \\
        \midrule
         
DA-MRG~\cite{peng2022domain} & p,c/nc,t & GraphSAGE & Combining multi-relational graph learning with domain-aware classifiers in a federated framework. \\
\hline
FedKG~\cite{wang2024fedkg} & p,c/nc,t & RGCN & Distilling knowledge across federated RGCNs using a generator-guided optimisation to reduce client-side data diversity. \\
\hline
FedACK~\cite{yang2023fedack} & p,c/nc,t & GCN & Reweighting neighbourhood influence in graph convolution using an attention mechanism and federated learning. \\

\bottomrule 
\end{tabularx}
\end{table*}

\subsection{Adversarial Attacks}
\label{subsec:DL-AA}

In graph-based DL, adversarial attacks involve small perturbations to features or the graph structure that can mislead models such as GCN. These attacks expose critical vulnerabilities in GNN for tasks like node classification, particularly in applications such as FAD, where adversaries may manipulate the graph’s structure or node attributes to evade detection~\cite{zugner2018adversarial}.
In the context of FAD, adversarial attacks target the victim model, which refers to the detection model deployed by an OSN platform. This model can be treated as a black box, where the attacker lacks access to its architecture, parameters, or training data, or as a white box, when such information is available. Regardless of the attacker's level of knowledge, the objective remains the same: to mislead the detection model into misclassifying fake accounts as real.

Liu et al.~\cite{liu2024social} introduced an adversarial attack approach by targeting node attributes. They proposed the Attribute Random Iteration Fast Gradient Sign Method (\textbf{ARI-FGSM}), an adversarial attack technique designed to evaluate the robustness of FAD models. This method perturbs the numerical attributes of accounts, such as the number of followers, friends, statuses, and active days, while keeping the underlying graph structure unchanged.
ARI-FGSM operates under both white-box and black-box settings. In white-box scenarios, it leverages gradient information from the target model. It uses a substitute model to generate transferable adversarial examples in black-box settings. The attack follows an iterative process: In each step, one numerical attribute is randomly selected and adjusted in the direction that increases the model’s classification loss. The process terminates early if the model misclassifies the fake account as real. To maintain realism and subtlety (minimal perturbations), all perturbed values are clipped within the original dataset’s minimum and maximum ranges.
To improve robustness, a simple training strategy is applied. Successful adversarial samples are collected during the attack phase and used to retrain the model, thereby strengthening its defences with low computational cost and no degradation in classification performance.

Wang et al.~\cite{wang2023my} proposed a black-box adversarial attack on FAD that departs from traditional attribute- or structure-based approaches by introducing a node injection–based strategy. The attack aims to inject a single, carefully crafted fake node adjacent to a target fake account, such that both evade detection by the victim model. This represents an attempt to explore node injection under black-box constraints in the context of FAD.
The framework consists of four key stages. First, a substitute model is trained using an RGCN to approximate the behaviour of the inaccessible victim model based on publicly available account attributes and graph structure. Then, an embedding for the injected node is generated using information from the target fake node, its neighbours, and class-specific transformation weights from the substitute model. 
Then, a single edge is added between the injected node and either the target node or one of its first-order neighbours, preserving stealth and limiting detectability through a slight modification to the original graph.
Finally, an attribute recovery module reconstructs realistic input-space features (e.g., numerical and categorical metadata) from the generated embedding. This step ensures the injected node appears plausible and practically deployable in real-world OSNs.

The two studies discussed in this section highlight distinct adversarial strategies in FAD. Structural attacks, such as node injection, expose the vulnerability of graph-based models to network structure manipulation under black-box settings. In contrast, fine-grained attribute perturbations demonstrate that minimal yet realistic changes to account metadata can significantly degrade detection performance, even without altering the graph topology.

\begin{table*}[ht!]
    \centering
    \caption{Adversarial attacks (Section~\ref{subsec:DL-AA}). Backbone: base DL models. Features: c, g, p, per (content, graph, profile, personal); nc, t (numerical/categorical, textual).} 
    % \label{tab:graph_ml_summary}
    %\renewcommand{\arraystretch}{1.2}
    \begin{tabularx}{\textwidth}{|>{\hsize=0.15\hsize}X |>{\hsize=0.12\hsize}X |>{\hsize=0.14\hsize}X |>{\hsize=0.59\hsize}X|}
        \toprule
        \textbf{Paper} & \textbf{Features} & \textbf{Backbone} & \textbf{Main Idea} \\
        \midrule
         
ARI-FGSM~\cite{liu2024social} & p,c,g/nc,t & GNNs Robustness Evaluation & Evaluating and improving the robustness of FAD via attribute perturbations. \\
\hline
Wang et al.~\cite{wang2023my} & p,c,g/nc,t & RGCN & Fooling FAD by injecting adversarial nodes with recovered attributes to hide targeted fake accounts in the social graph. \\
\bottomrule 
\end{tabularx}
\end{table*}

\section{Datasets}
\label{sec:dataset}

The datasets used for FAD range from fully synthesised graphs to real-world OSNs, annotated either manually or automatically. The nature and origin of a dataset significantly influence the generalizability, reproducibility, and validity of the proposed detection methods. While some prior works have extracted and annotated their own datasets from real-world OSNs (cf.~\cite{ercsahin2017twitter,balaanand2019enhanced}), others have relied on publicly available benchmark datasets or synthesised data. Synthesised datasets enable controlled experimentation under predefined assumptions, whereas real-world datasets capture complex social behaviours, offering a more realistic and challenging evaluation environment. More detailed discussions on both datasets are provided as real-world datasets (Section~\ref{subsec:dataset-benchmark}) and synthesised datasets (Section~\ref{subsec:dataset-synthesise}).

\subsection{Real-World Datasets}
\label{subsec:dataset-benchmark}

Benchmark datasets from real-world platforms such as Twitter and Facebook are essential for evaluating FAD models. These datasets typically include accounts labelled manually or semi-automatically, enabling supervised learning and performance assessment. labelling strategies commonly rely on rule-based heuristics~\cite{balaanand2019enhanced}, manual annotation~\cite{shi2023mgtab}, account suspension records~\cite{gao2018sybilfuse}, or the creation of fake accounts by the researchers themselves~\cite{cresci2015fame}.

Publicly available datasets promote reproducibility and comparability across studies. Some also provide diverse features beyond the graph structure, such as textual, profile, and behavioural data. Below, we cover the most popular datasets for FAD.

% The following datasets are among the most comprehensive, incorporate graph structures, and have been widely adopted in recent FAD research.

Cresci et al.~\cite{cresci2015fame} introduced a large-scale, publicly available dataset for FAD on Twitter, commonly referred to as \textbf{Cresci-15}. The dataset contains 1950 verified human accounts and 3351 fake accounts, balanced to support effective classifier training. Human accounts were collected from two sources: volunteers who followed a project account and passed CAPTCHA verification, and a manually labelled political dataset. Fake accounts were purchased from three online markets. This dataset provides shared tweets from accounts as content.
In addition to account-level labels, Cresci-15 includes graph structure, profile features, content features, and graph-based features. The social graph is modelled using directed edges representing the following relationships.

Feng et al.~\cite{feng2021twibot20} introduced \textbf{TwiBot-20}, a large-scale and comprehensive Twitter FAD benchmark designed to address limitations of previous datasets, such as low account diversity, limited account information, and insufficient data volume. TwiBot-20 includes over 229,000 Twitter accounts and provides three modalities of account information: textual content and profile features, numerical and categorical profile metadata, and graph-based features. The social graph is modelled using directed edges representing the following/follower relationships.
By encompassing a wide range of account interests (e.g., politics, business, entertainment, and sports), diverse geographic regions, and rich annotations, TwiBot-20 supports both account-level and community-aware FAD. The dataset was annotated through a crowdsourcing process in which five active Twitter accounts labelled each node. Annotators followed detailed guidelines based on prior research, which outlined five key characteristics of fake accounts, such as a lack of originality and high automation. Accounts verified by Twitter were automatically labelled as human. For other accounts, labels were assigned based on a consensus from at least four out of five annotators. If consensus was not achieved, the research team attempted to contact the accounts directly via Twitter and conducted manual reviews, discarding ambiguous cases.

Later, Feng et al.~\cite{feng2022twibot22} introduced \textbf{TwiBot-22}, a large-scale graph-based Twitter FAD benchmark that improves dataset size, label annotation quality, and feature diversity. TwiBot-22 represents the Twitter network as a multi-relational graph with four types of entities (accounts, tweets, lists, and hashtags) and 14 types of relations. These include account-to-account (follows, followed by), account-to-tweet (posts, likes, pins), tweet-to-tweet (mentions, retweets, quotes, replies to), account-to-list (owns, membership, follows), list-to-tweet (contains), and tweet-to-hashtag (discusses).
Compared to prior graph-based datasets such as Cresci-15 and TwiBot-20, which contain only account entities and tweets and three relation types (follows, followed by, and account-tweet post relations), TwiBot-22 offers a substantially richer and more realistic graph structure for Twitter-based FAD research.

Shi et al.~\cite{shi2023mgtab} introduced \textbf{MGTAB}, a large-scale, expert-annotated benchmark designed for both stance detection\footnote{Stance detection is the task of determining whether an account or a piece of content supports, opposes, or remains neutral toward a given target within a text.} and FAD on Twitter. Compared to previous benchmarks, MGTAB provides a significantly larger set of expertly labelled accounts, making it suitable for both supervised and semi-supervised learning settings. The dataset represents Twitter as a multi-relational account graph with seven types of relations: follower, following (referred to as a friend relation in this dataset), mention, reply, and quote (explicit relations), as well as hashtag and URL co-occurrence (implicit relations). In addition, tweets (textual content features) are encoded using LaBSE, a multilingual BERT-based model, and then combined with numerical, categorical, and textual profile features for downstream tasks. MGTAB contains $10199$ expert-annotated accounts and 400,000 additional unlabelled accounts provided in the \textbf{MGTAB-large} version.

\subsection{Synthesised Datasets}
\label{subsec:dataset-synthesise}

Although a variety of real-world OSN datasets are publicly available, such as those from the SNAP repository~\cite{snapnets}
% \footnote{\url{https://snap.stanford.edu/data/}} 
and the Network repository~\cite{rossi2015network}, labelling them for the FAD task remains a substantial challenge. This challenge arises due to privacy constraints, restricted access to internal platform information (such as account reports), and the difficulty in reliably determining ground-truth labels. 

Synthesised datasets are commonly constructed to evaluate the performance of detection methods under controlled and repeatable conditions. By simulating the behaviour of fake accounts, these datasets provide explicit ground-truth labels, which support a more precise analysis of a method’s theoretical capabilities. Despite not fully capturing the complexity and unpredictability of fake account behaviour observed in real-world OSNs, synthesised datasets offer considerable flexibility. They allow researchers to systematically manipulate structural and connectivity patterns in the graph, thereby enabling comprehensive evaluations across a range of adversarial attack scenarios~\cite{cao2012aiding}. In addition, some synthesising algorithms have been developed specifically to build upon unlabelled real-world datasets, retaining real-world information.  

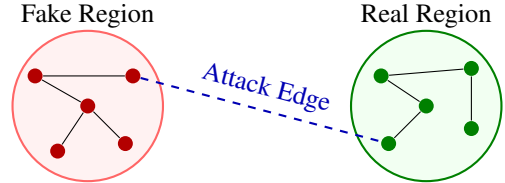
\begin{figure}[ht]
    \centering
    \begin{tikzpicture}[scale=1.0]
        % Styles
        \tikzset{
            fakecircle/.style={draw=red!60, thick, fill=red!5},
            realcircle/.style={draw=green!50!black, thick, fill=green!5},
            fakenode/.style={circle, fill=red!70!black, inner sep=2pt},
            realnode/.style={circle, fill=green!50!black, inner sep=2pt},
            attackedge/.style={thick, dashed, blue!70!black}
        }
        \def\R{1.0cm}
        % Centers of circles (closer together)
        \coordinate (FCenter) at (0,0);
        \coordinate (RCenter) at (4.5,0);
        % Draw regions
        \draw[fakecircle] (FCenter) circle (\R);
        \draw[realcircle] (RCenter) circle (\R);
        % Labels for regions
        \node at ($(FCenter)+(0,1.2)$) {Fake Region};
        \node at ($(RCenter)+(0,1.2)$) {Real Region};
        % Fake region nodes (5)
        
        \node[fakenode] (f1) at ($(FCenter)+(-0.7,0.4)$) {};
        \node[fakenode] (f2) at ($(FCenter)+(-0.4,-0.6)$) {};
        \node[fakenode] (f3) at ($(FCenter)+(0.5,-0.5)$) {};
        \node[fakenode] (f4) at ($(FCenter)+(0.6,0.4)$) {};
        \node[fakenode] (f5) at ($(FCenter)+(0,0)$) {};
        % Real region nodes (5)
        \node[realnode] (r1) at ($(RCenter)+(-0.6,0.4)$) {};
        \node[realnode] (r2) at ($(RCenter)+(-0.5,-0.5)$) {};
        \node[realnode] (r3) at ($(RCenter)+(0.6,-0.3)$) {};
        \node[realnode] (r4) at ($(RCenter)+(0.6,0.5)$) {};
        \node[realnode] (r5) at ($(RCenter)+(0,0)$) {};
        % Edges within fake region
        \draw (f1) -- (f5);
        \draw (f2) -- (f5);
        \draw (f3) -- (f5);
        \draw (f4) -- (f1);
        % Edges within real region
        \draw (r1) -- (r5);
        \draw (r2) -- (r5);
        \draw (r3) -- (r4);
        \draw (r4) -- (r1);
        % Attack edge
        \draw[attackedge] (f4) -- node[above, sloped, text=blue!70!black]{Attack Edge} (r2);
    \end{tikzpicture}
    \caption{A synthesised graph showing a Fake Region and a Real Region connected by an Attack Edge.}
    \label{fig:synthesizingdatasetgeneralfigure}
\end{figure}

To synthesise datasets for FAD, it is essential to define a strategy for constructing both the fake and real regions, including their respective nodes and the internal connections (either directed or undirected edges). Additionally, a strategy is required to determine how these two regions are connected. This connection typically occurs through attack edges, which link nodes in the fake region to those in the real region. In the case of directed graphs, an additional question is how to connect the real region to the fake region. Figure~\ref{fig:synthesizingdatasetgeneralfigure} illustrates an example of such a structure, showing the two regions and an attacking edge between them. In the following, we discuss various methods for specifying each of these components across different dataset synthesis techniques.

Cao et al.\cite{cao2012aiding} modelled OSNs as undirected graphs, where edges represent mutual relationships. They constructed their dataset by combining real-world unlabelled graphs (e.g., Facebook) with synthesised graph models. For the real region, they used either an unlabelled real graph or a graph generated by the Barabási–Albert (BA) model~\cite{barabasi1999emergence}. To model the fake region, they selected either the Erdős–Rényi (ER) model~\cite{erdos1960evolution} or the BA model. Two attack strategies were introduced for connecting the fake and real regions: a random attack strategy, where each fake node connects to randomly selected real nodes, and a targeted attack strategy, where fake nodes connect to real accounts near a predefined set of real nodes.

The use of ER and BA models to generate the internal structure of fake regions is common across several studies (cf.~\cite{zhang2022enhancing, wei2012sybildefender, lu2023sybilhp}). Some works, such as SybilHP~\cite{lu2023sybilhp} and Dehkordi and Zehmakan~\cite{dehkordi2025efficient}, further simulate fake regions by replicating parts of the real region’s structure and injecting attack edges based on their defined strategies.

Recently, the notion of account resistance was introduced to synthesise more dynamic and realistic datasets for FAD~\cite{dehkordi2025efficient}. In this framework, the fake region is constructed by replicating the structure of a selected portion of the real region, creating a corresponding fake node for each selected real account. Edges between fake and real accounts, referred to as attack edges, are formed through a request-acceptance simulation governed by resistance. Each real account is assigned a binary resistance label: resistant accounts reject incoming fake requests, while non-resistant accounts accept them.
The authors proposed three strategies for sending requests from fake to real accounts: a random strategy, where fake accounts send requests to real accounts uniformly at random; a modified preferential attachment strategy, which computes the probability like BA strategy but with higher weight for previously accepted attack edges from other fake accounts; and a BFS-based strategy, where each fake account targets the neighbours of its corresponding real dual node. Furthermore, with some probability, a real account that accepts a fake request may also initiate a reciprocal connection, thereby increasing the realism of the synthesised graph.

Table~\ref{table:dataset_stats} summarises key statistics of widely used datasets in FAD research. For each dataset, we also provide a sample list of representative papers that have employed them for evaluation in Table~\ref{tab:dataset_examples}.

\begin{table*}[ht!]
    \centering
    \caption{Summary of widely used Twitter-based FAD datasets with graph structure. \#$\mathfrak{X}$ denotes the number of elements in $\mathfrak{X}$, Accts stands for accounts. The last column stands for textual features such as tweets.}
    \label{table:dataset_stats}
    \begin{tabularx}{\textwidth}{|>{\hsize=0.18\hsize}X |>{\hsize=0.16\hsize}X |>{\hsize=0.16\hsize}X |>{\hsize=0.16\hsize}X|>{\hsize=0.16\hsize}X|>{\hsize=0.18\hsize}X|}
        \toprule
        \textbf{Dataset} & \textbf{\#Nodes} & \textbf{\#Fake Accts} & \textbf{\#Real Accts} & \textbf{\#Edges}  &  \textbf{Has Text?}  \\
        \midrule
         
Cresci-15~\cite{cresci2015fame}  & 5,301 &3,351&1,950&7,086,134&Yes\\
\hline 
TwiBot-20~\cite{feng2021twibot20}  & 229,580 &6,589 &5,237  & 33,716,171 &Yes\\
\hline 
TwiBot-22~\cite{feng2022twibot22} & 1,000,000&13,9943&860,057&  170,185,937 &Yes\\
\hline 
MGTAB~\cite{shi2023mgtab}  & 10,199 & 2,748&7,451&1,700,108&encoded as\hspace{2cm} numerical vector\\
\hline
MGTAB-large~\cite{shi2023mgtab}  &410,199&  2,748&7,451& 97,997,710 &encoded as\hspace{2cm} numerical vector\\
\bottomrule 
\end{tabularx} 
\end{table*}

\begin{table*}[ht!]
    \centering
    \caption{Examples of papers using each of the mentioned datasets.}
    \label{tab:dataset_examples} 
    \begin{tabularx}{\textwidth}{|>{\hsize=0.07\hsize}X |>{\hsize=0.93\hsize}X |}
        \toprule
        \textbf{Dataset} & \textbf{Papers} \\ 
        \midrule
         
Cresci-15 & \cite{chen2024cacl}, \cite{huang2024cgnn}, \cite{lei2022bic}, \cite{li2022sybilflyover}, \cite{li2023multi}, \cite{li2024botcl}, \cite{li2025ets}, \cite{liu2023Accou2Vec}, \cite{liu2023botmoe}, \cite{liu2024social}, \cite{peng2024coarse}, \cite{pham2022bot2vec}, \cite{qiao2023social}, \cite{qiao2024dispelling}, \cite{shi2023rf}, \cite{shi2024over}, \cite{shi2024sstackgnn}, \cite{wang2023my}, \cite{wang2024unsupervised}, \cite{yang2023rosgas}, \cite{ye2023hofa} \\
\hline
MGTAB & \cite{gong2024enhancing}, \cite{huang2024cgnn}, \cite{li2023multi}, \cite{miao2024bsg4bot}, \cite{qiao2024dispelling}, \cite{shi2023rf}, \cite{shi2024over}, \cite{shi2024sstackgnn}, \cite{xu2024dgbot}, \cite{yang2024sebot}, \cite{ye2023hofa} \\
\hline
TwiBot-20 & \cite{chen2024cacl}, \cite{du2024tcae}, \cite{feng2021botrgcn}, \cite{feng2022heterogeneity}, \cite{gong2024enhancing}, \cite{he2024dynamicity}, \cite{huang2024cgnn}, \cite{lei2022bic}, \cite{li2023multi}, \cite{li2024botcl}, \cite{li2025ets}, \cite{liu2023botmoe}, \cite{liu2024segcn}, \cite{lu2024adaptive}, \cite{miao2024bsg4bot}, \cite{peng2022domain}, \cite{peng2024coarse}, \cite{qiao2023social}, \cite{qiao2024dispelling}, \cite{reiche2024integrating}, \cite{shi2023rf}, \cite{shi2024over}, \cite{shi2024sstackgnn}, \cite{tzoumanekas2024graph}, \cite{wang2024botrga}, \cite{wang2024fedkg}, \cite{wang2024unsupervised}, \cite{wu2023botscl}, \cite{xu2024dgbot}, \cite{yang2024sebot}, \cite{ye2023hofa}, \cite{zhou2023semi} \\
\hline
Twibot-22 & \cite{alothali2022bot}, \cite{chang2024sega}, \cite{chen2024cacl}, \cite{he2024dynamicity}, \cite{li2024botcl}, \cite{liu2023botmoe}, \cite{liu2024segcn}, \cite{liu2024social}, \cite{miao2024bsg4bot}, \cite{reiche2024integrating}, \cite{wang2023my}, \cite{wang2024botrga}, \cite{wu2023botscl}, \cite{zhou2023detecting}, \cite{zhou2023semi} \\
\bottomrule 
\end{tabularx} 
\end{table*}

\section{Conclusion and Future Work}
\label{sec:conc+fut}
We reviewed a broad spectrum of FAD methods, with a focus on graph-based approaches. We discussed classical algorithms, traditional ML models, and recent DL models. Although the field has progressed from simple to sophisticated DL models, several challenges remain that point to valuable directions for future research. In this section, we first outline potential future work and then offer final concluding remarks.

\subsection{Future Work}

Despite notable progress in FAD, several research challenges and underexplored areas remain, opening up exciting future directions worth exploring. This subsection outlines key directions that could advance the field, including dataset development, heterophily handling, explainability, and adversarial robustness.

\paragraph{Dataset Development and Data-Centric Research.}

In the age of data-driven AI technologies, datasets play a crucial role. They are essential for training models and serve as the foundation for fair and meaningful evaluation. As discussed in Section~\ref{sec:dataset}, there is only a handful of benchmark datasets, and creating high-quality datasets remains an inherently challenging task for advancing research on FAD.

For instance, most existing benchmark datasets for FAD are from Twitter, but expanding to other platforms is essential. Each OSN has its unique structure and characteristics, and evaluating models across different platforms can lead to more reliable and generalisable results.
% In addition, as people often have accounts across multiple OSNs, building datasets that reflect this cross-platform behaviour can lead to more realistic FAD models.
Different platforms support different types of media. Some focus mainly on text, while others include images, audio, or video. Each media type can carry unique information and hidden patterns that FAD models can leverage. As a result, providing datasets with multiple modalities can be highly beneficial.

In this survey, the term``fake account'' is used in a general sense. Generally, FAD studies approach the detection task as a binary classification problem. However, providing a dataset with more detailed labels, such as spammers, robots, or data collectors, would be highly useful. This allows FAD to be framed as a multi-class classification problem, better capturing the diversity of fake account behaviours and leading to more accurate and robust models.

\paragraph{Focus on Heterophilic Edges.}
As mentioned, heterophilic edges cause feature mixing between fake and real accounts during the message passing process, resulting in blurred class boundaries and increased errors in node classification~\cite{wu2023botscl}.
In addressing the heterophilic edges challenge in FAD, a growing and promising line of research involves incorporating causality between nodes, which entails identifying cause-effect dependencies by estimating asymmetric relationships between node pairs, thereby enabling the detection of heterophilic edges that often degrade GNN performance~\cite{wang2025heterophilic}.
While several studies have explored causality from various perspectives, key questions remain: How should causality between nodes be defined and computed? Moreover, how can it be effectively incorporated into node classification tasks? Although multiple valid formalisms for causality exist~\cite{lin2022orphicx}, their integration into graph learning remains a challenge. While causality-based methods are gaining attention in general graph learning, tailoring these approaches specifically for FAD may uncover novel opportunities.
Wang et al.~\cite{wang2025heterophilic} took a step in this direction by investigating causal relationships between connected nodes, using this insight to identify heterophilic edges. However, many aspects of using causality remain open to debate. For example, how can causality be adequately defined and computed in graphs with multiple edge types? Addressing such questions could lead to more principled and effective strategies for mitigating the impact of heterophilic edges in FAD.

\paragraph{Opportunities for Few-Shot, Zero-Shot.}
While labelling nodes is often costly and time-consuming, the potential of few-shot and zero-shot learning for node classification has been well explored in recent literature, cf.~\cite{chen2024review}. 
However, they have not yet been fully explored or adapted for FAD, where such approaches could offer significant advantages in low-supervision settings. More focus on integrating these learning paradigms into FAD frameworks could open new paths for improving detection performance with minimal labelled data.

\paragraph{Cold-Start in FAD.}
Another promising direction lies in addressing cold-start nodes, a challenge that has been extensively studied in other domains, such as recommendation systems (e.g.~\cite{chen2025graph}). Cold-start nodes are those new nodes with limited connections, making them difficult to classify using traditional graph-based methods~\cite{jacobs2024g}. This challenge aligns closely with the early detection of FAD, as accounts are often poorly connected at the beginning. 

\paragraph{Explainability.}
It is challenging to fully trust the FAD methods without understanding how they make predictions. This lack of transparency (especially in DL models) limits their application in practical domains where fairness, privacy, and safety are key concerns. To confidently and responsibly deploy models in such settings, it is essential to ensure that the models are accurate and make their decisions understandable to humans~\cite{yuan2022explainability}.
Similarly, it is essential to understand why a model flags a particular account as fake in the context of FAD, which helps build trust in the system and supports its responsible use. Moreover, explainability enables the identification of potential issues, such as when a model relies on biased or misleading patterns in its decision-making process.

\paragraph{Adversarial Attacks.}
While a few studies have investigated adversarial attacks in the context of FAD~\cite{wang2023my,liu2024social}, our understanding of adequate adversarial attacks remains limited, and further research is needed. Designing realistic and sophisticated attack strategies that resemble actual fake account behaviours observed in real-world networks is essential. These strategies can also serve as benchmarks for evaluating the robustness of FAD models.

\subsection{Conclusion}

This survey provided a comprehensive overview of FAD methods in OSNs, with a strong focus on graph-based approaches. By systematically analysing the structural properties of social graphs and incorporating extra features, including profile features, content features, and graph-based features, we presented a coherent overview of how various models aim to detect real and fake accounts.
We discussed classical algorithmic methods, such as RW and LBP solutions. Then, we progressed toward ML and DL methods that rely on graph-based representations, such as label propagation, representation learning, and contrastive training, which leverage the graph's topology to enhance classification accuracy. The review also considered key structural challenges, such as heterophilic edges in networks, multiple types of edges in graphs, and graphs with multiple node types.
In addition, we examined how research in this area addresses adversarial attacks, limited label availability, and evolving attack strategies. We highlighted the importance of learning paradigms that can adapt to dynamic, sparse, and partially labelled environments.

This survey is a comprehensive and structured reference that synthesises existing research, identifies core challenges, and compares the diverse algorithmic and learning-based strategies in this domain. We hope that this work will support the development of more robust, interpretable, and generalisable FAD systems in real-world OSNs, making them safer.

\bibliographystyle{IEEEtran} 
\bibliography{references}

\end{document}